\documentclass[copyright,creativecommons]{eptcs}
\providecommand{\event}{SYNT 2015} 
\usepackage{amsthm}
\usepackage{amssymb}
\usepackage{mathtools}
\usepackage{temporal}
\usepackage{enumerate} 
\usepackage{color}
\usepackage{tikz}
\usepackage{paralist}
\usepackage{fixme,manfnt,mparhack}
\usepackage[T1]{fontenc}
\usepackage{enumerate}
\usepackage{xspace}
\usepackage{amssymb}
\usepackage{paralist}
\usepackage{wrapfig}
\usepackage{bbm}
\usepackage{cite}

\usepackage{macros}

\providecommand{\thisvolume}[1]{this volume of {\sl Electronic
  Proceedings in Theoretical Computer Science}}

\definecolor{mygreen}{rgb}{0,0.6,0} 
\definecolor{mygray}{rgb}{0.9,0.9,0.9} 
\definecolor{mymauve}{rgb}{0.58,0,0.82}

\usepackage{listings}
\usepackage[margins]{trackchanges}
\addeditor{SJ}

\usepackage{xspace}

\newcommand{\mysubsubsection}[1]{\medskip \noindent {\bf #1.}}
\newcommand{\myparagraph}[1]{\smallskip \noindent {\em #1.}}

\newcommand{\NN}{\mathbb{N}}

\newcommand{\depqbf}{\textsf{DepQBF}\xspace}
\newcommand{\bloqqer}{\textsf{Bloqqer}\xspace}
\newcommand{\demiurge}{\textsf{Demiurge}\xspace}

\newcommand{\syntcomp}{SYNTCOMP\xspace}
\newcommand{\minisat}{\textsf{Minisat}\xspace}
\newcommand{\lingeling}{\textsf{Lingeling}\xspace}

\newcommand{\abssynthe}{AbsSynthe\xspace}
\newcommand{\simpleBDD}{Simple BDD Solver\xspace}
\newcommand{\realizer}{Realizer\xspace}
\newcommand{\picosat}{\textsf{PicoSat}\xspace}

\newcommand{\qbfcert}{\textsf{QBFCert}\xspace}

\newcommand{\Abc}{\textsf{ABC}\xspace}
\newcommand{\aiger}{\textsf{AIGER}\xspace}

\newcommand{\subby}{\kern-0.2em\leftarrow\kern-0.2em}

\title{The Second Reactive Synthesis Competition\\ (\syntcomp 2015)}
\author{Swen Jacobs
\institute{Saarland University\\Saarbr\"ucken, Germany}
\and
Roderick Bloem
\institute{Graz University of Technology \\ Graz, Austria}
\and 
Romain Brenguier
\institute{Universit\'e Libre de Bruxelles\\ Brussels, Belgium}
\and 
Robert K\"onighofer
\institute{Graz University of Technology \\ Graz, Austria}
\and
Guillermo A. P\'erez 
\institute{Universit\'e Libre de Bruxelles\\ Brussels, Belgium}
\and 
Jean-Fran\c{c}ois Raskin
\institute{Universit\'e Libre de Bruxelles\\ Brussels, Belgium}
\and
Leonid Ryzhyk
\institute{
\begin{tabular}{cc}
NICTA & Carnegie Mellon University\\
Sydney, Australia & Pittsburgh, USA
\end{tabular}}
\and 
Ocan Sankur 
\institute{Universit\'e Libre de Bruxelles\\ Brussels, Belgium}
\and 
Martina Seidl
\institute{Johannes-Kepler-University\\ Linz, Austria}
\and 
Leander Tentrup
\institute{Saarland University\\Saarbr\"ucken, Germany}
\and 
Adam Walker
\institute{NICTA \\ Sydney, Australia}
}
\def\titlerunning{\syntcomp 2015}
\def\authorrunning{S. Jacobs et al.}
\begin{document}
\maketitle

\begin{abstract}
We report on the design and results of the second reactive synthesis competition (\syntcomp 2015). We describe our extended benchmark library, with $6$ completely new sets of benchmarks, and
additional challenging instances for $4$ of the benchmark sets that were already
used in \syntcomp 2014. To enhance
the analysis of experimental results, we introduce an extension of our benchmark
format with meta-information, including a difficulty
rating and a reference size for solutions.
Tools are evaluated on a set of $250$ benchmarks, selected to provide a good
coverage of benchmarks from all classes and difficulties.
We report on changes of the evaluation scheme and the experimental setup.
Finally, we describe the entrants into \syntcomp 2015, as well as the results of our
experimental evaluation. In our analysis, we emphasize progress over the tools
that participated last year.
\end{abstract}

\section{Introduction}
\label{sec:intro}

The automatic synthesis of reactive systems from formal specifications is one of the major challenges of computer science. While there has been lively research on the topic for more than 50 years and
a number of fundamental approaches to solve the problem have been proposed~\cite{Church62,EmersonC82,Rabin69,PnueliR89}, the impact of theoretical results on the practice of system design has been rather limited. This is in spite of the obvious advantages of the automatic construction of provably correct systems and an increased interest in possible applications of reactive synthesis techniques, e.g., in robotics and cyber-physical systems.\cite{ChinchaliLTBM12,Kress-GazitFP09,ChenDSB12} 
The goal of the reactive synthesis competition (\syntcomp) is to increase the impact of theoretical advancements in synthesis by fostering research in scalable and user-friendly implementations of synthesis
techniques.

In particular, \syntcomp aims at
\begin{enumerate}[i)]
\itemsep0pt
\item making synthesis tools comparable by establishing a standard benchmark format,
\item facilitating the exchange of benchmarks by providing a public benchmark repository and encouraging researchers to contribute their benchmarks,
\item enabling a comprehensive and fair evaluation of synthesis tools by providing a \emph{dedicated and independent} platform for the comparison of tools under consistent experimental conditions,
\item encouraging the implementation of synthesis tools that can be used as \emph{black-box solvers} in applications by enforcing that competition entrants have to run on the complete benchmark set of \syntcomp with a fixed configuration, and
\item fostering the efficient implementation of synthesis algorithms by regularly providing new and challenging benchmark problems, and comparing the performance of tools on these.
\end{enumerate}


The first \syntcomp was held during June and July of 2014, and its results were
first presented at the 26th International Conference on Computer Aided
Verification (CAV) and the 3rd Workshop on Synthesis (SYNT) in July 2014.~\cite{Jacobs15}
A design choice of the first competition was to focus on safety properties
  specified as monitor circuits in an extension of the AIGER format (known from the hardware model checking competition~\cite{HWMCC}). We keep this restriction for the second competition, and plan to consider extensions of the format in the future.

The organization team of \syntcomp 2015 consisted of Roderick Bloem and Swen Jacobs.
Based on results and experiences from the first competition, the specific goals for the second competition were: 
\begin{itemize}
\itemsep0pt
\item to expand the benchmark library with new and challenging problems,
\item to improve the selection of benchmarks to ensure an even distribution over benchmarks from different classes and difficulties,
\item to improve the ranking scheme, focusing on the most important properties of synthesis tools, and
\item to determine the progress of the state of the art by comparing the performance of new entrants to those from last year.
\end{itemize}

The rest of this paper describes the design, benchmarks, participants, and
results of \syntcomp 2015. We describe the synthesis problem as considered in \syntcomp in Section~\ref{sec:problem}, followed by a presentation of the benchmark set for \syntcomp 2015 in Section~\ref{sec:benchmarks}. 
Section~\ref{sec:setup} introduces the rules and setup of the competition, including classification and selection of benchmarks. In Section~\ref{sec:participants} we give an overview of the entrants of \syntcomp 2015, focusing on changes with respect to last year's entrants. Finally, we present and analyze the experimental results in Section~\ref{sec:results}.

Note that Sections~\ref{sec:benchmarks} and~\ref{sec:participants} are in part based on descriptions supplied by the respective authors of benchmarks and tools. The
remainder of this article is original work of the \syntcomp organizers.

\section{Synthesis Problem}
\label{sec:problem}

We briefly summarize the reactive synthesis problem as it is considered in \syntcomp. A more detailed introduction into the problem, as well as into the different approaches for solving it, can be found in~\cite{Jacobs15}.

In \syntcomp, we consider the automatic synthesis of finite-state reactive systems that satisfy a safety specification, given as a monitor circuit that raises a special output \out when an unsafe state is visited. The specification is encoded in the \syntcomp format, an extension of the well-known AIGER format~\cite{aiger} that allows inputs of the circuit to be defined as either controllable or uncontrollable. In the traditional game-based approach to the synthesis of reactive systems~\cite{BL69,Rabin69,Thomas95}, such a specification gives rise to a \emph{game }between
two players: states of the game are given by the valuation of latches in the monitor circuit, the environment player decides on the uncontrollable inputs of the specification circuit, and the system player decides on the controllable inputs. The goal of the system player is to satisfy
the specification, i.e., to visit only safe states, independent of the environment
behavior.  

Algorithms that solve this game usually take an approach that consists of two steps. In the first step, a so-called \emph{winning region} is computed. The
winning region $W$ is the set of all states from which the system player can enforce
to satisfy the specification, i.e., to visit only safe states in the subsequent computation. In the classical algorithm, this is done by computing the fixpoint of the \emph{uncontrollable predecessor} operation \upre on the error states, i.e., inductively computing all states from which the environment can force the game into the unsafe states. Since two-player safety games are determined, the complement of this set is the winning region $W$ of the system player.
In a second step, a \emph{winning strategy} is derived from the
winning region.  For every (current) state and uncontrollable input, the winning strategy
defines a set of controllable inputs that are okay for satisfying the specification. In order to obtain an implementation of this strategy as a circuit, a concrete choice
for the controllable inputs has to be made for every state and uncontrollable input.

In order to achieve acceptable scalability, it is important to implement
synthesis algorithms symbolically, i.e., by manipulating formulas instead of
enumerating states. In synthesis, these symbolic algorithms are usually implemented with Binary Decision Diagrams (BDDs)~\cite{bryant86,somenzi99,AlurMN05}.  One reason for this is
that solving games inherently involves dealing with quantifier alternations,
and BDDs offer techniques for handling both kinds of quantification. However, BDDs also have their
scalability issues.  
On the other hand, there have been enormous performance
improvements in decision procedures for the satisfiability of (boolean) formulas over the
last years and decades.  This has lead to efficient tools like SAT- and QBF
(Quantified Boolean Formulas) solvers, which can also be leveraged to obtain efficient symbolic synthesis algorithms.

All of the tools that compete in \syntcomp 2015 implement symbolic game-based synthesis in some form. A description of the tools will be given in Section~\ref{sec:participants}.

\section{Benchmarks}
\label{sec:benchmarks}

In this section, we describe the benchmark library for \syntcomp 2015. We start by describing in detail $6$ new classes of benchmarks. This is followed by short descriptions of the classes of benchmarks that have already been used in \syntcomp 2014 --- for some of these, additional problem instances have been added, and two of them are split into multiple classes for \syntcomp 2015.

\subsection{Scheduling of Washing Cycles (new class)}

This class of benchmarks specifies a washing system, with cycles
that can be launched in parallel. The system is
composed of several tanks that may depend on shared water pipes, and can be controlled by the user pushing
buttons.  It is parameterized in the number of tanks $n$, the maximum allowed reaction delay $d$, as well as the number $t$ of tanks that share a water pipe.
The correct behavior of the system is described by rational
expressions that impose safety constraints.  These safety constraints are first 
translated to non-deterministic automata, and then into the \syntcomp format. The benchmark set is described in more detail in~\cite{BrenguierPRS15}.

\begin{figure}[ht]
  \centering{
    \begin{tikzpicture}[>=stealth',shorten >=1pt,shorten <=1pt,auto,node
		    distance=2cm,every loop/.style={looseness=6},initial
		    text={},every fit/.style={draw,densely
		    dotted,rectangle},el/.style={font=\footnotesize}]
      \draw (0,0) node[draw,thick,minimum width=2cm,minimum height=2cm] (WT)
      {Water tank $i$};
      \draw[thick,-latex'] (-2.5,.5) |- (WT.180);
      \draw (-2.5,.5) node[left,above] {$\textsf{fill}_i$};
      \draw[-latex',thick] (WT.0) -- +(0.8,0) -- +(0.8,-0.5) node[below]
      {$\textsf{empty}_i$};

      \draw (4.5,.5) node[draw,minimum width=0.8cm,minimum height=0.3cm] (B) {};
      \draw (4.5,-.5) node (Push) {$\textsf{push}_i$};
      \draw[-latex'] (B) -- (Push);
      \draw (B.-90) +(1,0) node (E)  {};
      \draw (B.-90) +(-1,0) -- (E);
    \end{tikzpicture}
  }
  \caption{A water tank.  Input $\textsf{fill}_i$ controls the arrival of
  water into the tank. Input $\textsf{empty}_i$ controls the departure of water out of the tank.  Input $\textsf{push}_i$ initiates the washing cycle when the button is pushed.}
  \label{fig:tank}
\end{figure}
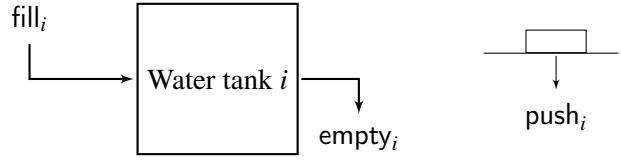

\paragraph{Problem description.}
We consider the design of a centralized controller for a washing system, composed of $n$ tanks running in parallel.
The user can request the delivery of water to a given tank $i$ by pressing a button, modeled as uncontrollable input $\textsf{push}_i$. The release of water into the tank is modeled by controllable input $\textsf{fill}_i$. After a determined time $k$, the controller should open the valve to remove the water from inside the tank, modeled by controllable input $\textsf{empty}_i$.  Finally, there is a controllable input $\textsf{light}$ that models an activation light that should be on whenever one of the tanks contain water. This is illustrated
in Fig.~\ref{fig:tank}.

\paragraph{Encoding of the problem by safety specification.}
We encode the specification for the controllers into rational expressions
that denote \emph{the occurrence of an error} in the execution: 

\def\true{\textsf{true}}
\begin{itemize}
\item
An error occurs if a button was pressed but the tank is not filled after
delay $d\in \mathbb{N}$:
\[
A_i = \true^* \cdot \{\textsf{push}_i\} \cdot \{ \lnot \textsf{fill}_i \}^d.
\]

\item
The tank should not be filled unless the button was pushed:
\[
B_i = \true^* \cdot \{\lnot \textsf{push}_i\}^d \cdot \{ \textsf{fill}_i \}.
\]
\item 
The tank should be emptied after a delay of exactly $k$ steps:
\begin{align*}
  C_i &= \true^* \cdot \{\textsf{fill}_i\} \cdot \true^k \cdot \{ \lnot
	  \textsf{empty}_i \} \\
  C'_i &= \true^* \cdot \{\textsf{fill}_i\} \cdot (\true \mid \varepsilon)^{k-1}
  \{ \textsf{empty}_i \} 
\end{align*}

\item
The light should be on if, and only if, one of the $n$ tanks is being
filled:
\[
D =  \true^* \cdot \left\{\textsf{light} \ne \bigvee_{i \in [1,n]}
\textsf{fill}_i \right\}.
\]

\item
If tanks $i$ and $j$ are connected to the same pipe, as illustrated in
Fig.~\ref{fig:pipes}, then the system should not activate $\textsf{fill}_i$ and $\textsf{fill}_j$ at the same time. We write $P$ for the set of pipes, and for a pipe~$p\in
P$, we write $i \in p$ to denote that tank~$i$ is fed by pipe~$p$.  We
encode the mutual exclusion constraint by raising an error for the following
expression:
$$
E = \true^*\cdot \left\{ \bigvee_{p \in P} \bigwedge_{i,j \in p} \textsf{fill}_i
\right\}
$$
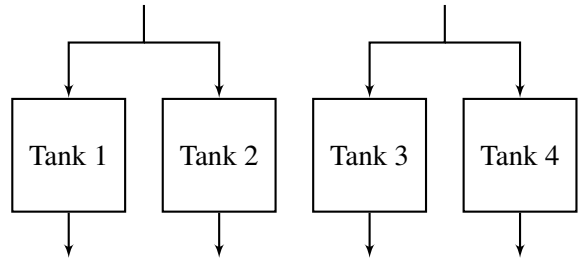
\begin{figure}[ht]
  \centering{
    \begin{tikzpicture}
      \draw (0,0) node[draw,thick,minimum width=1.5cm,minimum height=1.5cm]
      (WT1) {Tank $1$};
      \draw (2,0) node[draw,thick,minimum width=1.5cm,minimum height=1.5cm]
      (WT2) {Tank $2$};
      \draw (4,0) node[draw,thick,minimum width=1.5cm,minimum height=1.5cm]
      (WT3) {Tank $3$};
      \draw (6,0) node[draw,thick,minimum width=1.5cm,minimum height=1.5cm]
      (WT4) {Tank $4$};
      \draw[thick,-latex'] (1,2) -- (1,1.5) -| (WT1.90);
      \draw[thick,-latex'] (1,1.5) -| (WT2.90);
      \draw[thick,-latex'] (5,2) -- (5,1.5) -| (WT3.90);
      \draw[thick,-latex'] (5,1.5) -| (WT4.90);
      \draw[-latex',thick] (WT1.-90) -- +(0,-0.6);
      \draw[-latex',thick] (WT2.-90) -- +(0,-0.6);
      \draw[-latex',thick] (WT3.-90) -- +(0,-0.6);
      \draw[-latex',thick] (WT4.-90) -- +(0,-0.6);
    \end{tikzpicture}
  }
  \caption{A system with 4 water tanks. 
    Tanks $1$ and $2$ connect to the same pipe, and so do tanks $3$ and $4$.}
  \label{fig:pipes}
\end{figure}
\end{itemize}

Overall, we consider the safety specification given by the language:
\[
\mathcal{L} = \bigcup_{i\in [1,n]} \left( L(A_i) \cup L(B_i) \cup L(C_i) \cup
L(C'_i)\right) \cup L(D) \cup L(E)
\]
Where $L(A)$ denotes the languages associated to expresssion $A$. 

\paragraph{Translation to AIGER.}
To produce benchmarks in AIGER format, the conjunction of rational
expressions is first translated to a non-deterministic automaton~$\mathcal{A}$. 
Each state of $\mathcal{A}$ is encoded by a latch in the AIGER file. In an
execution of the circuit, a latch~$l$ is on if, and only if, there is an
execution of $\mathcal{A}$ --- on the prefix of the word that has been read so
far --- that leads to the state corresponding to latch~$l$.  The error latch is
put to true if, and only if, one of the latches corresponding to accepting
states is on.
 The error
output will be set to true whenever the current prefix belongs
to the language $\mathcal{L}$. 

In the benchmark package, the AIGER files are named
\texttt{cycle\_sched\_}$n$\_$d$\_$t$\texttt{.aag}, where $n$ is the number of
tanks, $d$ is the maximum delay between $\textsf{push}_i$ and $\textsf{fill}_i$,
and $t$ is the number of tanks that are alimented by the
same pipes. The delay $k$ between $\textsf{fill}_i$ and
$\textsf{empty}_i$ is always chosen to be equal to $d$.

This benchmark set contains $321$ instances, and has been supplied by Romain Brenguier. The OCaml program that has been used to generate AIGER files from rationnal
expressions can be downloaded from the following
address:~\url{https://github.com/romainbrenguier/Speculoos}.

\subsection{Driver Synthesis (new class)}
This class of benchmarks is based on a driver synthesis benchmark for the Termite synthesis tool~\cite{RyzhykWKLRSV14,termite}. 
In Termite, the driver synthesis problem is encoded as a game played by the driver (or controller) against its environment, consisting of an operating system and the device to be controlled. The given benchmark considers the synthesis of a driver for an IDE (or PATA) hard disk controller. The driver must perform three basic functions: read sectors, write sectors and device configuration.

\paragraph{Problem description.}
The benchmark consists of two interacting state machines:
\begin{itemize}
    \item an operating system (OS), and
    \item an IDE hard disk controller. 
\end{itemize}

The OS acts as a workload generator for generating possible OS-driver interactions. In this benchmark, the specification requires
\begin{itemize}
\item that the driver both read and write hard disk sectors, 
\item that these sectors are eventually read/written, and
\item that the driver does not perform erroneous actions such as reading/writing hard disk sectors when not requested. 
\end{itemize}
The requirements that the OS specification imposes are expressed as reachability goals, e.g., any outstanding hard disk write is eventually performed.

The IDE hard disk controller state machine models the external register-mapped control interface of this device as well as an internal state machine that performs the disk sector reads and writes. By writing the correct values to these registers in the correct order, the driver can initiate read and write transactions in order to  satisfy the requirements imposed by the OS workload generator. Most registers contain data such as the sector offset to start writing, number of sectors to write, and a pointer to the data's location in memory. Others contain control bits, which, when written, trigger various operations in the hardware such as a disk write. 

The goal is to synthesize a driver that provides values for the control bits such that the specification is satisfied. Conceptually, the driver has full information of the system state and when it detects that there is an outstanding request, it performs the necessary device register reads and writes to force the IDE controller state machine to perform a read or write, thus resolving the request.

\paragraph{Parameterization and encoding into AIGER.}
The control bits of the hard disk controller are modeled as controllable inputs. 
The uncontrollable inputs model non-determinism in the OS and device state machines. This ensures that the driver can respond correctly to all possible generated workloads.

Benchmarks from this class are named \texttt{driver\_XYZ.aag}, where \texttt{X}, \texttt{Y} and \texttt{Z} stand for the three dimensions in which the benchmark is parameterized:
\begin{itemize}
\item the specification contains a number of state variables that represent data that is inconsequential to the property to be synthesized. One dimension of parameterization is an abstraction of such state variables, by either removing them completely, or reducing the bit-width of certain signals. $\texttt{X=a}$ stands for the variant of the benchmark without data abstraction, and $\texttt{X=b,c,d}$ are three variants that abstract increasingly more of the data.
\item the original specification contains a liveness constraint, which is reduced to a safety property by requiring the desired property to hold at least once within every execution fragment of some fixed length, given by $\texttt{Y}$. \texttt{Y} ranges between $2$ and $10$, where small values result in unrealizability of the benchmark, while large values increase the size of the state space and potentially make it more difficult to find a solution.
\item finally, \texttt{Z} can take two values, where $\texttt{Z=n}$ means that the benchmark has not been further modified, and $\texttt{Z=y}$ means that \Abc\cite{abc} has been used to simplify the circuit.
\end{itemize}

This benchmark was supplied by Adam Walker, and translated from Verilog into AIGER by Robert K\"onighofer, using the VL2MV routine of the VIS system~\cite{VIS}, followed by a translation to AIGER format (and possibly optimization) by \Abc. The source code and exact sequence of commands are replicated in the comments section of each benchmark file. Overall, this class contains $72$ problem instances.

\subsection{Huffman Encoder (new class)}
This class of benchmarks is taken from recent work of Khalimov~\cite{Khalimov15} on a framework for specifying synthesis problems. The idea is: given a Huffman decoder~\cite{Huffman52}, synthesize the encoder.

The specification includes a given Huffman decoder, which reads the outputs of the encoder that is to be synthesized, and a monitor circuit that compares environment inputs to the encoder with the outputs of the decoder --- they should match. Furthermore, the encoder is required to eventually supply the encoded signals that correspond to its input. The benchmark is parameterized in the safety parameter $k$ that is used to approximate this liveness condition, i.e., we require the encoder to supply the encoded signals within $k$ steps. For the given decoder, the specification is unrealizable if $k\leq9$, and realizable otherwise. 

This benchmark was written in an extended version of the SMV format~\cite{SMVSystem,SMVLang}, and converted to AIGER by first translating it into standard, flattened SMV format, and then using \texttt{smvtoaig} and a liveness-to-safety approximation to obtain a file in the \syntcomp format. For more details on the benchmark, the original specification language and its translation into \syntcomp format, we refer to Khalimov~\cite{Khalimov15}.

\subsection{HWMCC (new class)}
This class of benchmarks is based on a subset of the benchmarks from HWMCC 2014~\cite{HWMCC}. The idea is to consider verification benchmarks consisting of a system and a safety specification that is not satisfied by the system, and ask the question whether we can synthesize a controller for a given subset of the inputs such that the specification is satisfied. 

To obtain the benchmarks in this set, we consider the unsafe instances from the \emph{single safety property} track of HWMCC. Depending on the overall number of inputs of the original benchmark, we consider variants where between $1$ and $512$ are defined to be controllable. The files retain their original filenames, with \texttt{\_c0toc}x appended to the filename if inputs $0$ to $x$ are declared as controllable. E.g., the benchmark based on \texttt{6s210b037.aag}, where $32$ inputs have been declared as controllable, is named {6s210b037\_c0to31.aag}.

This benchmark set contains $110$ benchmarks, including some of the largest AIGER files that have been considered in the competition so far, with file size of up to 6MB, corresponding to more than $20000$ inputs, $40000$ latches, and $200000$ AND-gates. The original verification benchmarks have been taken from the HWMCC website~\cite{HWMCC} and modified by Swen Jacobs.

\subsection{HyperLTL (new class)}
This class of benchmarks is based on a number of benchmark problems from HyperLTL model checking, as introduced in recent work by Finkbeiner et al.~\cite{FinkbeinerRS15}. HyperLTL allows to express properties of multiple executions of the same system, e.g., information-flow properties or symmetry properties. In this case, we consider given implementations of several mutual exclusion protocols inspired by the bakery protocol and the AMBA specification, and symmetry properties like $$\forall \pi, \pi'.\ (select(0)_\pi \land select(1)_{\pi'}) \rightarrow \always (pc(0)_\pi=pc(1)_{\pi'} \land pc(1)_\pi=pc(0)_{\pi'},$$
stating that if in an execution $\pi$ the scheduler has initially selected process $0$, and in execution $\pi'$ it has selected process $1$, then over the whole execution the program counters of these processes will be swapped.

The set of benchmarks contains different implementations of bakery protocols and the AMBA protocol, with different symmetry properties.

\paragraph{Encoding.} 
The HyperLTL model checking problems are translated to QBF queries, which are in turn encoded in AIGER format. The idea is that the synthesizer checks the HyperLTL property, and constructs a witness formula (represented as a controller circuit) if this is the case. 

The benchmark set contains $21$ benchmarks, and has been supplied by Markus Rabe. For more details on HyperLTL model checking and the benchmarks, we refer to Finkbeiner et al.~\cite{FinkbeinerRS15}.

\subsection{Matrix Multiplication (new class)}
These benchmarks describe the synthesis of a matrix multiplication circuit. That is, the inputs represent two boolean matrices, and the controller needs to give outputs that represent the product matrix of the inputs (in every step of the system).
Matrix multiplication is a basic operation that has many applications in
mathematics, physics, and engineering.  Implementing this operation with a
logical circuit of small size is important to produce hardware at small cost.
Furthermore, this is an example of a \emph{compositional} benchmark, as every entry of the output matrix only depends on one column of one input matrix and one row of the other, but not on the rest of the input matrices.

In addition to the basic matrix multiplication benchmarks, the benchmark set includes problems with repeated matrix multiplication, where a boolean matrix is stored in the circuit, and multiplied repeatedly with matrices defined by the current inputs. Half of the inputs are controllable, and the goal is to never obtain a matrix that has a line composed only of $0$s or only of $1$s.

\paragraph{Encoding.}
We consider the set of
Booleans $\mathbb{B} = \{ 0, 1 \}$ and multiplication of matrices in the Boolean
ring $\langle \mathbb{B}, \lor, \land, 0 , 1 \rangle$.  The inputs of the specification circuit encode matrices $A \in \mathbb{B}^{m \times n}$, $B \in \mathbb{B}^{n \times o}$, and
$C \in \mathbb{B}^{m \times o}$, with the inputs for matrix $C$ defined as controllable. The error output is set to true if, in any step, $A \cdot B \ne C$.
In the benchmark package, the AIGER file
\texttt{mult\_bool\_matrix\_}$m$\_$n$\_$o$\texttt{.aag} encodes the
mutiplication of a matrix $A$ of dimension $m \times n$ and a matrix $B$ of
dimension $n \times o$.

For the variant with repeated matrix multiplication, the inputs of the circuit encode a matrix $A \in \mathbb{B}^{n \times n}$ with $n \in \NN$, where inputs corresponding to the first $n/2$ columns are controllable. Furthermore, the circuit stores a matrix $B \in \mathbb{B}^{m \times n}$ with $m \in \NN$, initialized to a matrix that alternates between $0$ and $1$ (i.e., $B_{ij}=1$ iff $i+j$ is even), and updated in each step to the result of $A \cdot B$. The error output is raised if at any point $B$ is such that either $\exists j.\ \forall i.\ B_{ij}=0$, or $\exists j.\ \forall i.\ B_{ij}=1$.
In the benchmark package, the AIGER file
\texttt{mult\_bool\_matrix\_dyn\_}$m$\_$n$\texttt{.aag} encodes the repeated
mutiplication of a matrix $B$ of dimension $m \times n$ with an input matrix $A$ of
dimension $n \times n$.

This benchmark set was supplied by Romain Brenguier and contains $354$ problem instances --- $273$ for basic matrix multiplication, and $81$ for the repeated matrix multiplication variant. The AIGER files for these benchmarks have been generated from an OCaml program that can be downloaded from the following
address:~\url{https://github.com/romainbrenguier/Speculoos}.

\subsection{AMBA (extended class)}
The synthesis of a bus controller for the AMBA specification (see~\cite{BloemJPPS12}) was already considered as a benchmark last year~\cite{Jacobs15}. The benchmarks have been encoded in Verilog and then translated into AIGER, and are parameterized in three dimensions:
\begin{itemize}
\item The number of masters $n$ that want to access the bus.
\item The type of liveness-to-safety approximation (three different types).
\item A parameter $k$ for the liveness-to-safety approximation, requesting for example that a liveness property has to be satisfied after every $k$ steps of the whole system.
\end{itemize}
This year, this benchmark set has been extended by considering larger values for $n$ (up to $16$) and $k$ (up to $60$). Overall, the set now contains $952$ instances. 
This benchmark set has been supplied by Robert K\"onighofer and is described in more detail in~\cite{Jacobs15}.

\subsection{Genbuf (extended class)}
This benchmark set considers the synthesis of a generalized buffer (as described in~\cite{BloemJPPS12}), and is parameterized in the same way as the AMBA benchmarks above. This year, we consider values of up to $64$ for $n$, and $120$ for $k$. This benchmark set now contains $866$ instances, supplied by Robert K\"onighofer, and is described in more detail in~\cite{Jacobs15}.

\subsection{LTL2AIG (extended class)}
This set of benchmarks is based on the benchmark set of the Acacia synthesis tool~\cite{bbfjr12,acacia}, translated from specifications in LTL using the LTL2AIG tool~\cite{Jacobs15}. In 2014, only a part of the original benchmark set was translated, as the LTL2AIG tool did not support all features of the original specifications. This year, the complete benchmark set was entered into the competition. It now contains:
\begin{itemize}
\item $50$ instances of \texttt{demo} problems that are based on the test suite of LTL synthesis tool \textsc{Lily}\cite{JobstmannB06}, with specifications of traffic lights and arbiters in different complexity,
\item $41$ instances of \texttt{ltl2dba} and \texttt{ltl2dpa} problems that use the synthesis tool to obtain a deterministic B\"uchi automaton (dba) or a deterministic parity automaton (dpa) that corresponds to a given LTL formula,
\item $42$ instances of the \texttt{gb} benchmark, based on the same generalized buffer case study as the benchmark above, but in this case as a direct translation of an LTL specification into \syntcomp format, and
\item $64$ instances of a load balancer benchmark, originally presented with LTL synthesis tool \textsc{Unbeast}~\cite{Ehlers11}.
\end{itemize}
This benchmark set has been supplied by Guillermo A. P\'erez. The first two sets have already been present last year, the other two have been newly generated based on the new version of the LTL2AIG tool (and replace a small set of instances for the \texttt{gb} and load balancer benchmarks that have been present last year). Instead of a single benchmark class, we now consider this set as $4$ different benchmark classes in our selection of benchmarks shown in Table~\ref{tab:selected-benchmarks}.

\subsection{Toy Examples (extended class)}
This benchmark set considers a number of basic building blocks of circuits, such as an adder (\texttt{add}), a bitshifter (\texttt{bs}), a counter (\texttt{count}), or a multiplier (\texttt{mult}). All of the benchmarks are parameterized in the bitwidth of the input. This year, the benchmark set includes
\begin{itemize}
\item $50$ instances of benchmark \texttt{add},
\item $18$ instances of benchmark \texttt{bs},
\item $28$ instances of benchmark \texttt{cnt},
\item $14$ instances of benchmark \texttt{mult},
\item $42$ instances of benchmarks \texttt{mv} and \texttt{mvs}, and
\item $24$ instances of benchmark \texttt{stay}.
\end{itemize}
For benchmarks \texttt{add} and \texttt{bs}, instances with larger bit-width have been added this year. This benchmark set has been supplied by Robert K\"onighofer and is described in more detail in~\cite{Jacobs15}.

\subsection{Factory Assembly Line and Moving Obstacle (unchanged)}
These two benchmark sets describe a controller for two robot arms on an assembly line, and a moving robot that should avoid a moving obstacle in two-dimensional space, respectively. The factory assembly line benchmark consists of $15$ problem instances, and the moving obstacle benchmark of $16$. Both sets of benchmarks have been supplied by R\"udiger Ehlers and have already been used in \syntcomp 2014~\cite{Jacobs15}.

\section{Setup, Rules and Execution}
\label{sec:setup}

\subsection{Classification and Selection of Benchmarks}
\label{sec:selection}

To facilitate the selection of suitable benchmarks and the evaluation of experimental results, we collected additional information on realizability and difficulty of benchmark problems. For realizable specifications, we additionally determined the smallest known solution, to be stored as a reference size. For benchmarks that were already used in \syntcomp 2014, this information was obtained from the experimental results of the previous year. For the new benchmarks, we conducted \emph{classification experiments} by running the three most successful solvers from \syntcomp 2014 in sequential realizability mode on a representative subset of the new benchmarks.

\paragraph{Extension of the Benchmark Format.}
The data obtained in the classification is included directly in each benchmark file, as a special paragraph of the comments section. This paragraph starts with a line containing only the \syntcomp tag ``\texttt{\#!SYNTCOMP}'', and ends with a line containing only ``\texttt{\#.}''. Between these lines, properties of the benchmark can be defined. The properties defined for \syntcomp 2015 are listed in Table~\ref{tab:classification}.

Both \texttt{SOLVED\_IN} and \texttt{REF\_SIZE} may be set to $0$ if the problem has not been solved before. An example of a classification paragraph is given in Listing~\ref{list:classification}.

\begin{table}[b]
\renewcommand{\arraystretch}{1.5}
\caption{Properties defined for classification of benchmarks in \syntcomp 2015}
\label{tab:classification}
\begin{tabular}{ll}
Property & Value \\
\hline
\texttt{STATUS} & \texttt{realizable}, \texttt{unrealizable} or \texttt{unknown}\\
\texttt{SOLVED\_BY} & \parbox[t]{13cm}{the fraction of participants that solved the benchmark in a previous experiment, e.g., \texttt{8/8}, followed by a description of the experiment in brackets, e.g., \texttt{[SYNTCOMP2014-RealSeq]}}\\
\texttt{SOLVED\_IN} & \parbox[t]{13cm}{the minimal time (in seconds) needed to solve the problem in a previous experiment, followed by a description of the experiment in brackets}\\
\texttt{REF\_SIZE} & \parbox[t]{13cm}{the minimal size of a solution (in number of AND gates) produced in a previous experiment}
\end{tabular}
\end{table}

\lstset{caption={Classification paragraph of benchmark file \texttt{add8y.aag}},label=list:classification,captionpos=t}
\begin{lstlisting}[float]
	#!SYNTCOMP
	STATUS : realizable
	SOLVED_BY : 8/8 [SYNTCOMP2014-RealSeq]
	SOLVED_IN : 0.008 [SYNTCOMP2014-RealSeq]
	REF_SIZE : 203
	#.
\end{lstlisting}

\paragraph{Selection of Competition Benchmarks.}
To ensure a fair and meaningful evaluation of the participants, we want to ensure that benchmarks from different classes of problems have approximately equal weight, and that the competition benchmarks represent a good distribution across different difficulties for each class. To this end, benchmarks are divided into categories that can be either small or large. From each category, we selected a number of problems (usually $16$ for large categories, and $8$ for small) with an even distribution over difficulties, in terms of the ratio of solvers that were able to solve the benchmark previously, given in the \texttt{SOLVED\_BY} tag of the benchmarks. In particular, the selected benchmarks from every set include a fraction of about $20\%$ of benchmarks that have not been solved before. 
The number of selected problems from each category (cp. Section~\ref{sec:benchmarks}) is given in Table~\ref{tab:selected-benchmarks}.

\begin{table}[h]
\caption{Number of selected Benchmarks per Category}
\label{tab:selected-benchmarks}
\centering
\def\arraystretch{1.2}
\begin{tabular}{ll|ll}
Category & Benchmarks & Category & Benchmarks\\
\hline
AMBA & 16 & Add (Toy Examples)& 8\\
(Washing) Cycle Scheduling & 15 & Bitshift (Toy Examples)& 8\\
Demo (LTL2AIG)& 16 & Count (Toy Examples)& 8\\
Driver Synthesis & 16 & Genbuf (LTL2AIG) & 8\\
Factory Assembly Line & 15 & Huffman Encoder & 5\\
Genbuf & 16 & Mult (Toy Examples)& 8\\
HWMCC & 16 & Mv/Mvs (Toy Examples)& 8\\
HyperLTL & 15 & Stay (Toy Examples)& 8\\
Load Balancer (LTL2AIG)& 16\\
LTL2DBA/LTL2DPA (LTL2AIG)& 16\\
Moving Obstacle & 16 & \\
Matrix Multiplication & 16 & {\bf Total:} & {\bf 250}\\
\end{tabular}
\end{table}

\subsection{General Rules}
\label{sec:rules}
Like in the previous year, \syntcomp is divided into two main tracks: \emph{realizability checking }and \emph{synthesis}, as well as into two execution modes: \emph{sequential} and \emph{parallel}. We explain the rules of the competition, including evaluation of tools in the different tracks.

\paragraph{Input and Output Format.}
Participants have to accept input in the \syntcomp format, without further user intervention except for a fixed sequence of command-line parameters. In the synthesis track, solutions for realizable specifications also have to be provided in the \syntcomp format. Syntactic conformance to the format is automatically checked by a script provided by the organizers. For a description of input and output format, we refer to the report of the first competition~\cite{Jacobs15}. 

\paragraph{Timeout.}
In the sequential execution mode, the timeout for each problem is $3600$s of CPU time. In the parallel mode, the timeout is $3600$s of wall time.

\paragraph{Basic Ranking Scheme.}
In both tracks, a correct answer is rewarded with one point for the solver, and a wrong answer is punished by subtracting $4$ points.
Since most of the benchmarks were available to the participants before the competition and we allowed re-submission in case of implementation bugs that were detected in the testing phase of the competition, such a punishment was not necessary.
The tracks differ in how a solution is determined to be correct, as explained below.

\paragraph{Realizability Track.}
In this track, tools read the problem description and have to return \texttt{unrealizable} in case of an unrealizable specification, and \texttt{realizable} in case of a realizable specification. Correctness is determined based on the \texttt{STATUS} information in the classification paragraph of the benchmark, unless \texttt{STATUS} is \texttt{unknown}. In the latter case, correctness is determined by a majority vote of all solvers that provide a solution to the benchmark, and the execution platform for the experiments (see Section~\ref{sec:execution}) generates a notification that a previously unsolved problem has been solved.

\paragraph{Synthesis Track.}
In this track, tools read the problem description and have to return \texttt{unrealizable} in case of an unrealizable specification, and a solution that satisfies the specification and confirms to the \syntcomp format in case of a realizable specification. In addition to the information in the \texttt{STATUS} tag of the file, correctness of a solution is checked by running a model checker on the output file, with a separate timeout of 3600s. Only solutions that can be verified by the model checker are accepted. In case of problems that have status \texttt{unknown} and are solved for the first time, in this track there is \emph{no} majority vote on the correctness of the solution: if at least one solver produces a correct solution, it is assumed to be realizable. Similarly, if a problem is tagged with \texttt{unrealizable} and a solver produces a solution that is accepted by the model checker, we assume that the tag was wrong and the solution is correct.

\paragraph{Quality Ranking.} 
In addition to the basic ranking scheme, we define a \emph{quality ranking} for the synthesis track, in which solutions of realizable problems are awarded a different number of points, depending on the size of the solution $size_{new}$ and a reference size $size_{ref}$. The number of points for a solution is 
$$2 - \log_{10} \frac{size_{new}}{size_{ref}}.$$
That is, a solution that is of size $size_{ref}$ gets $2$ points; a solution that is bigger by a factor of $10$ gets $1$ point; a solution that is bigger by a factor of $100$ (or more) gets $0$ points; and similarly for solutions that are smaller than $size_{ref}$, e.g., a solution that is smaller by a factor of $10$ gets $3$ points.

For benchmark problems that have a \texttt{REF\_SIZE} tag (that is not equal to $0$), the value given there is taken as $size_{ref}$.
For benchmark problems that do not have such a tag because they are new or have not been solved in the last competition, we use the smallest size of any of the solutions of this year as the reference size.

\subsection{Execution}
\label{sec:execution}
\syntcomp 2015 used a set of identical machines with a single quad-core intel Xeon processor (E3-1271 v3, 3.6GHz) and 32 GB RAM (PC1600, ECC), running a GNU/Linux system. Each node has a local 480 GB SSD that can be used as temporary storage.

As in 2014, the competition was organized on the EDACC 
platform~\cite{BalintDGGKR11}, with a very similar setup.
To ensure a high comparability and reproducability of our results, a complete machine
was reserved for each job, i.e., one synthesis tool (configuration) running 
one benchmark. Olivier Roussel's 
\texttt{runsolver}~\cite{Roussel11}
was used to run each job and to measure CPU time and wall time, as well as 
enforcing timeouts. As all nodes are 
identical and no other tasks were run in parallel, no other limits than a 
timeout of $3600$ seconds (CPU time in sequential mode, wall time in
parallel mode) per benchmark was set.
Like last year, we used wrapper scripts to execute solvers that did not conform completely with the output format specified by the competition, e.g., to filter extra information that was displayed in addition to 
the specified output.

The model checker used for checking correctness of solutions is IIMC~\cite{IIMC} in version 2.0.

\section{Participants}
\label{sec:participants}
Four participants were entered into \syntcomp 2015. All of them follow the traditional game-based approach to the synthesis of reactive systems from safety specifications, as described in Section~\ref{sec:problem}.
 In this section, we briefly describe the methods implemented in each tool. Since all entrants have already participated in the competition in 2014, we focus on the changes when compared to last year's versions. For the BDD-based tools, we give an overview of the implemented methods in Table~\ref{tab:optimizations}. For detailed explanations of the different optimizations, we refer to last year's report~\cite{Jacobs15}.

\begin{table*}[h]
\caption{Optimizations implemented in BDD-based Tools.}
\label{tab:optimizations}
\centering
\def\arraystretch{1.1}
\begin{tabular}{r|ccccc}
  Technique                       & \abssynthe & \realizer & \simpleBDD \\  \hline
automatic reordering                        & x          & x         & x \\
eager dereferencing of BDDs       &            &           & x \\
direct substitution               & x          & x         & x \\
partitioned transition relation   & x                & x         & x \\
simultaneous conjunction and abstraction & &  & x \\
compositional synthesis  					& x					& & \\
abstraction-refinement						& (x)				& & x \\
co-factor based extraction of winning strategies & x & N/A      & N/A\\
forward reachability analysis     & x          & N/A       & N/A \\
\Abc minimization                 &            & N/A       & N/A\\
additional optimizations (see tool descriptions)         & x   & x         & x \\
\end{tabular}
\end{table*}

\subsection{AbsSynthe 2.0: compositional algorithms for synthesis}

\abssynthe was submitted by R. Brenguier, G. A. P\'erez, J.-F. Raskin, and 
O. Sankur from Universit\'e Libre de Bruxelles. \abssynthe implements 
different BDD-based synthesis approaches, and competed in both the realizability and the synthesis track.

\subsubsection{Overview}
The new version of AbsSynthe implements different BDD-based synthesis
algorithms, with and without decomposition into independent sub-games,
described in more detail in~\cite{BrenguierPRS15}. All algorithms use the BDD
package CUDD, with automatic BDD reordering using the sifting heuristic. In
sequential mode, three different algorithms were entered into the competition:
configuration seq1 uses a standard BDD-based fixpoint computation, with partitioned
transition relation and several other optimizations (see
Table~\ref{tab:optimizations}), but without compositionality. The other two configurations seq2 and seq3 use
different forms of compositional reasoning, explained below. An algorithm that uses abstraction
refinement is also implemented in AbsSynthe since its first version. However,
this abstraction-based algorithm did not compete this year. In addition, two parallel configurations par1 and par2 entered the competition, also explained below.

\subsubsection{Decomposing the Specification}\label{sec:decomp}
Let us demonstrate how we decompose the error function~$\fbad$
of a given symbolic game into a disjunction, i.e., $\fbad = \bigvee_i e_i$.
Notice that if a strategy (\ie~a controller) 
ensures that the error signal
is never true then it also ensures that $e_i$
is never true.
%
The rationale behind this approach is that the functions~$e_i$ do not depend on
all latches in general, so solving the game for~$e_i$ is often efficient.
%


Consider an AIG representing the formula
\[
	x_1 \land \lnot \left( x_2 \land \left( \lnot x_3 \land x_4 \right)
	\right),
\]
where $x_1,x_2,x_3,x_4$ are all input variables.
We can rewrite the formula as follows
\(
	\phi_{1} \equiv x_1 \land \lnot \phi_{2}
\)
where $\phi_{2}$ is
\(
	\phi_{2} \equiv x_2 \land \lnot x_3 \land x_4.
\)
If we distribute the disjunction from $\lnot \phi_{2}$ we get that
\(
	\phi_{1} \equiv (x_1 \land \lnot x_2) \lor
		(x_1 \land x_3) \lor 
		(x_1 \land \lnot x_4).
\)
Thus, one possible decomposition of $\phi_{1}$ would be to take $e_1 = x \land
\lnot x_2$, $e_2 = x_1 \land x_3$, and $e_3 = x_1 \land \lnot x_4$.

These general steps can be generalized into an algorithm which outputs a
decomposition of the error function whenever one exists. Intuitively, the
algorithm consists in exploring all non-inverted edges of the AIG graph from the
vertex which defines the error function. If there are no inverted edges which
stopped the exploration, or if all of them lead to leaves, the error function is
in fact a conjunction of Boolean variables and can clearly not be decomposed.
Otherwise, there is at least one inverted edge leading to a node representing an
AND gate.  In this case, we can push the negation one level down and obtain a
disjunction which can be distributed to obtain our decomposition. 

Given the set of $e_i$
\begin{inparaenum}[$(i)$]
\item we first simplify the transition relation using the classical
	\emph{generalized co-factor} BDD operation and keep it precise only in
	for latch transition functions which affect $e_i$.
\item Next, we solve each sub-game and obtain the set of states from which a
	winning strategy exists for the controller.
\item When combining solved sub-games, we further modify the transition relation
	by making every transition not allowed by winning strategies of the
	controller in sub-game go to an error state.
\end{inparaenum}

\subsubsection{Compositional Algorithms}\label{sec:algos}
We provide three different algorithms that first solve the sub-games
corresponding to the sub-circuits obtained by our decomposition procedure, and
then aggregate, following three different heuristics, the results obtained on
the sub-games. Namely, once we have the solutions of all the sub-games we
aggregate them by using one of the three heuristics described below.

\paragraph{Global aggregation.}
We start by computing the intersection of the
winning valuations of all sub-games: $\Lambda = \bigwedge_{1\leq i \leq n}
w_i(L,X_u,X_c)$, where each $w_i(L,X_u,X_c)$ is a winning valuation of the latches $L$, uncontrollable inputs $X_u$, and controllable inputs $X_c$, for subgame $i$. In fact, any valuation that is not in~$\Lambda$ is losing in
one of the sub-games; thus in the global game. Conversely, a strategy that stays
in~$\Lambda$ is winning for each sub-game.  Therefore, we solve the global game
with the new safety objective of avoiding $\lnot \Lambda$.  Before solving the
global game, the algorithm attempts to reduce the size of the transition
relations using \emph{restrict}. This algorithm entered \syntcomp 2015 as configuration seq2.

\paragraph{Incremental aggregation.}
In this algorithm, we aggregate the results of the sub-games
\emph{incrementally}: given the list of winning valuations $w_i$ for the
sub-games, at each iteration, we choose and remove two sub-games~$i$ and~$j$,
solve their conjunction (as with global aggregation, with error function
$\lnot (w_i \land w_j)$), and add the newly obtained winning valuations back in
the list. The choice of the pair $i,j$ is done heuristically. This algorithm entered \syntcomp 2015 as configuration seq3.

\paragraph{Back-and-forth.}
For this last algorithm, we interleave the analysis of the global game
(with objective $\Lambda$) and the analysis of the sub-games. Let $u(L)$ be the
set of ``bad'' states (initially containing all states where $\out$ is true). At
each iteration, we extend the losing states $u(L)$ by one step, by applying once
the $\upre$ operator in the global space.  We then consider each sub-game, and
check whether the new set $u'(L)$ of losing states (projected on the sub-game),
changes the local winning states. For those games in which it does grow,
we recompute the local fixed point.
We update the strategies~$\lambda_i$ of the
sub-games when necessary, and restart until stabilization.  Because analyzing
the sub-games is often more efficient than analyzing the global game, this
algorithm improves over the global aggregation algorithm in some cases (see the
experiments' section).  A similar idea was used in \cite{fjr10} for the problem
of synthesis from LTL specifications. This algorithm did not enter \syntcomp 2015 as a separate configuration in sequential mode, but was used as one thread in one of the parallel modes, see below.

\paragraph{Parallel Algorithms.}
Two parallel versions of \abssynthe entered the competition. Configuration par1 launches four threads in parallel, which execute the algorithms of sequential configurations seq1, seq2 and seq3, plus another thread with back-and-forth compositional reasoning. Configuration par2 also launches four threads in parallel, but in this case with four instances of strategy seq1 that only differ in the reordering strategy for BDDs.

\subsubsection{Strategy Extraction}
Strategy extraction in \abssynthe uses the co-factor-based approach described in~\cite{BloemGJPPW07}, with some additional optimizations as described in~\cite{Jacobs15}.

\subsubsection{Experiments}\label{sec:experiments}
Amongst the $674$ benchmarks considered in~\cite{BrenguierPRS15} (which include those
used in \syntcomp $2014$), $351$ are decomposable by our
static analysis into at least $2$ smaller sub-games. More specifically, the
average number of sub-games our decomposition algorithm outputs is $29$; the
median is $21$.

In general, the performances of the three compositional algorithms can differ,
but they are complementary.  
Even if AIG synthesis problems are monolithic, the
experiments show that our compositional approach was able to solve problems that
can not be handled by the monolithic backward algorithm; our compositional
algorithms are sometimes much more efficient. There are also examples that can
be decomposed but which are not solved more efficiently by the compositional
algorithms. So, it is often a good idea to apply all the algorithms in parallel.
%

\subsubsection{Implementation, Availability}
\abssynthe is implemented in C++, and depends only on a simple AIG
library (fetched from~\cite{aiger}) and the CUDD binary decision diagram
package~\cite{somenzi99}. The source code is available at
\url{https://github.com/gaperez64/AbsSynthe/tree/native-dev-par}.
More details on \abssynthe can be found in~\cite{BrenguierPRS14,BrenguierPRS15}.

\subsection{\demiurge 1.2.0: A SAT-Based Synthesis Tool}

\demiurge was submitted by R. K\"onighofer from Graz University of 
Technology and M. Seidl from Johannes-Kepler-University Linz. \demiurge 
implements several symbolic synthesis algorithms based on SAT and QBF solvers.
\demiurge competed in both the realizability and the synthesis track.

\subsubsection{Overview}
\begin{figure}[tb]
  \begin{center}
    \includegraphics[width=0.4\textwidth]{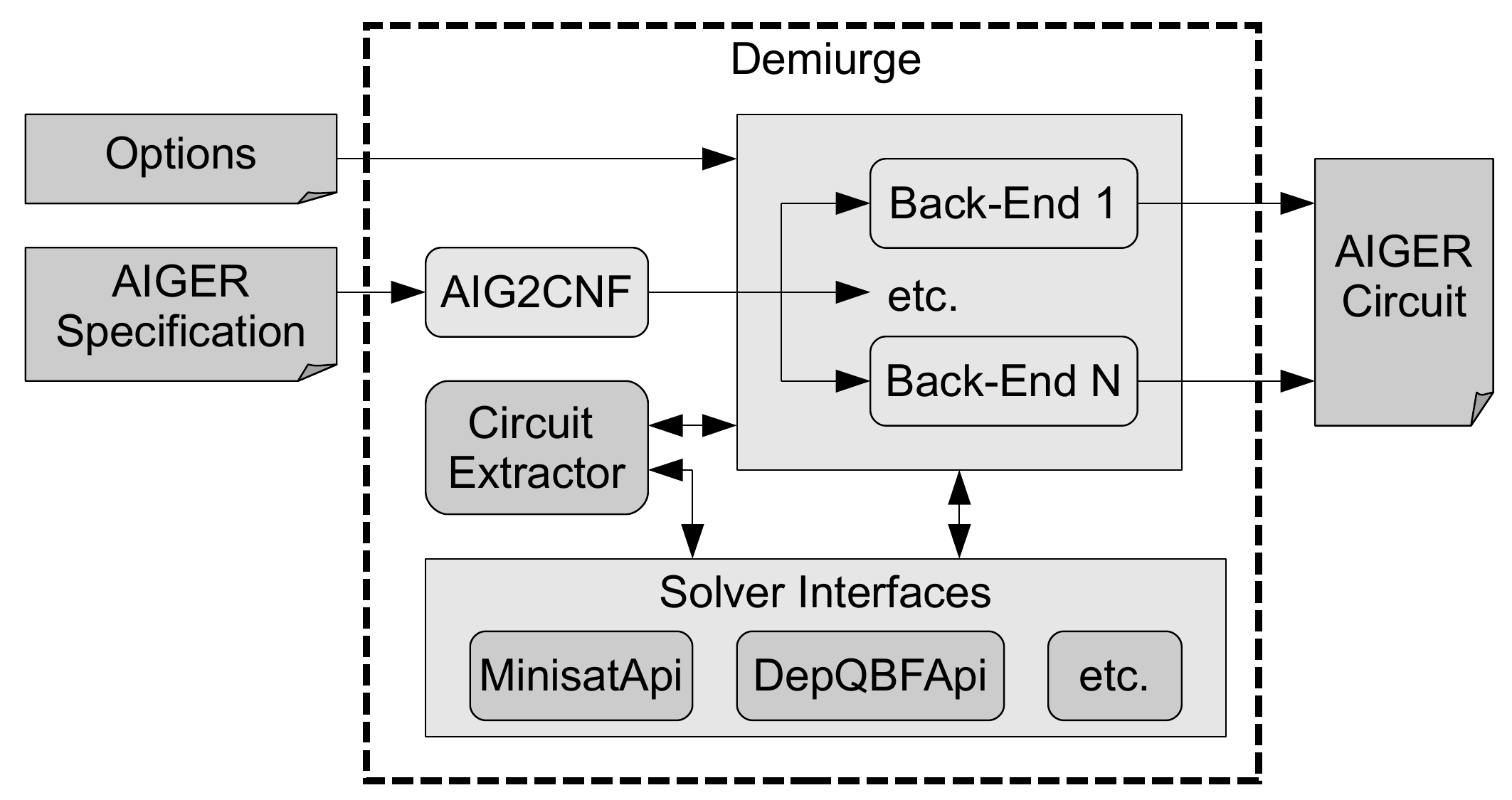}
    \caption{The architecture of \demiurge.}
    \vspace*{-0.3cm}
    \label{fig:arch}
  \end{center}
\end{figure}

The architecture of \demiurge is outlined in Fig.~\ref{fig:arch}.  The input is 
a safety specification in \aiger format.  The \textsf{AIG2CNF} module parses it 
into CNF formulas representing the transition relation and the set of safe 
states.  Next, the back-end selected by the user is executed.  The back-ends 
mostly differ in their method for computing the winning region, and can be 
parameterized with a method for computing a circuit from the winning region. 
Both the computation of the winning region and the extraction of circuits rely 
on external solvers like SAT- and QBF solvers.  The resulting circuits are 
optimized with \Abc~\cite{abc} and dumped in \aiger format again.

\subsubsection{Back-Ends}

\paragraph{Learning-Based Back-End.}

The learning-based back-end computes a CNF representation of the winning region
$W$ in an iterative manner.  It starts with the set of all safe states. In each
iteration, it computes a state within the current version $F$ of the winning
region from which the environment can enforce to leave $F$. Obviously, such a
state cannot be part of the final winning region $W$. Hence, the algorithm
refines $F$ by removing this state.  The state is represented as a cube over the
state variables, so removing it from $F$ amounts to adding a clause. By dropping
literals from the cube as long as it only contains states that must be
excluded from the winning region, the algorithm generalizes the state into a
larger region before removing it from the winning region.  The detailed
algorithm can be found in~\cite{BloemKS14}. 

For \syntcomp 2015, we use the following configuration, called D1synt if it is called with circuit extraction, and D1real otherwise.  Instead of a QBF 
solver, we use two competing SAT solvers to compute and generalize states to be 
removed from the winning region (algorithm \textsc{LearnSat} 
from~\cite{BloemKS14} with optimization \textsc{RG} enabled, but optimization 
\textsc{RC} disabled).  As a difference to the \syntcomp 2014 submission 
(version 1.1.0), we also apply partial universal quantifier expansion to reduce 
the number of iterations.  \minisat version 2.2.0 is the underlying SAT solver.

\paragraph{Template-Based Back-End.}
In order to obtain a winning region, this back-end constructs a parameterized 
CNF formula over the state variables: different concrete values for the 
(Boolean) parameters induce a different concrete CNF formula over the state 
variables.  This way, the search for a formula over the state variables (the 
winning region) is reduced to a search for Boolean constants (the template 
parameter values)~\cite{BloemKS14}.  While \demiurge 1.1.0 could only use a QBF 
solver for finding a template instantiation, version 1.2.0 can also use a SAT 
solver in a Counterexample-Guided Inductive Synthesis (CEGIS) approach.
For \syntcomp 2015, this back-end is not run separately, but only as one thread 
in our parallelization.

\paragraph{Incremental Induction Back-End.}
The \emph{incremental induction back-end} is a re-implementation of~\cite{MorgensternGS13}, generalizing IC3-based reachability checking~\cite{Bradley11} to the synthesis case.
For \syntcomp 2015, this back-end is not run separately, but only as one thread 
in our parallelization.

\paragraph{Parallel Back-End.}
The parallel back-end is a playground for combining different methods that 
refine a CNF representation of the winning region iteratively with additional 
clauses.  Several threads compute and add additional clauses in parallel.

For \syntcomp 2015, we use configurations with 3 threads, called P3synt or P3real for the synthesis and realizability tracks, respectively:  One thread executes the learning-based 
back-end, one the template-based back-end (alternating between QBF- and SAT 
solving in $20$ second turns), and one our incremental induction back-end.
\minisat 2.2.0 is used as SAT solver.  \depqbf 3.04  
with our extension of the QBF preprocessor \bloqqer~\cite{SeidlK14} is used for 
QBF solving in the template-based thread. Note that our parallelization is not 
just a portfolio approach. The different threads share clauses of the winning 
region as soon as they are discovered such that other threads can immediately 
benefit from this information.

\subsubsection{Circuit Extraction}

\demiurge provides several methods for computing circuits from the winning 
region~\cite{BloemEKKL14}.  One uses \qbfcert to compute Skolem functions for 
the output signals in a QBF that asserts completeness of the strategy relation.  
The second one uses computational learning to compute circuits for one output 
after the other.  A third method is based on interpolation.

For \syntcomp 2015, we use the learning approach (method \textsf{SL} 
from~\cite{BloemEKKL14}) with \lingeling \textsf{ayv} as SAT solver.  In our 
parallelization, we use 3 threads.  The first two execute the learning approach 
in two variants (\textsf{SL} and \textsf{SLN} from~\cite{BloemEKKL14}) .  The 
third thread executes the learning approach using incremental QBF solving 
(method \textsf{QL} from~\cite{BloemEKKL14})  with \depqbf 3.04.

\subsubsection{Implementation, Availability}
\demiurge is implemented in C++, and depends on a number of underlying
reasoning engines. Currently, \demiurge contains uniform 
interfaces to the APIs of \minisat, \lingeling, \picosat, and \depqbf (with and 
without the QBF preprocessor \bloqqer~\cite{SeidlK14}).  The interface to 
\depqbf also supports incremental QBF solving~\cite{LonsingE14}. Interfaces to 
SAT and QBF solvers supporting the \textsf{(Q)DIMACS} format are available as 
well. Furthermore, \demiurge interfaces \Abc~\cite{abc} for circuit 
minimization.
The source code is
available at {\url{%
http://www.iaik.tugraz.at/content/research/design_verification/demiurge/}}
under the GNU Lesser General Public License version 3.  The downloadable archive 
also contains extensive experimental results on the \syntcomp 2014 benchmarks 
and scripts to reproduce them.

\subsection{\realizer}
\label{sec:realizer}
\realizer was submitted by L. Tentrup from Saarland University, 
Saarbr\"ucken. \realizer implements BDD-based realizability checking, and
competed in the realizability track.
It does not support extraction of strategies.
We only give a very brief description of \realizer, since there are only minor changes to the version that competed in \syntcomp 2014~\cite{Jacobs15}.

\subsubsection{Synthesis algorithms}
\realizer implements the standard BDD-based fixpoint algorithm for safety games. 
It is based on BDD package CUDD, and uses automatic reordering of BDDs
with the \emph{lazy sift} reordering scheme. 
The fix-point algorithm is implemented in two variants, differing only in
the way they handle the transition relation of the circuit: one variant uses a
monolithic transition relation, while the other uses a partitioned transition
relation. Only the variant with partitioned transition relation competed in sequential execution mode, as the other one is not competitive in general. A number of additional optimizations are implemented in \realizer, compare Table~\ref{tab:optimizations}.

The main difference (besides bug fixes) to the version that competed in 2014 is the \emph{parallel mode}, which uses both variants of the algorithm running
(independently) in parallel.

\subsubsection{Implementation}
\realizer is written in Python and uses the BDD library CUDD in version 2.4.2
with the corresponding Python bindings PyCUDD in version 2.0.2.

\subsection{Simple BDD Solver 2}
Simple BDD Solver was submitted by L. Ryzhyk from NICTA, Sydney and the 
Carnegie Mellon University, Pittsburgh, and A. Walker from NICTA, Sydney. Simple
BDD Solver implements BDD-based realizability checking, and
only competed in the realizability subtrack.
It does not support extraction of strategies.

\subsubsection{Overview}
Simple BDD Solver is a substantial simplification of the solver that was
developed for the Termite project (\path{http://termite2.org}), adapted to safety
games given in the AIGER format.
It uses the BDD package CUDD, with dynamic variable reordering using the sifting
algorithm~\cite{Rudell93}, and a number of additional optimizations (again, cp. Table~\ref{tab:optimizations} and \cite{Jacobs15}). The basic algorithm entered the competition as configuration $1$, and is almost identical to the version that competed last year.

Furthermore, the tool implements a variant of the fixpoint algorithm with an
abstraction-refinement loop inspired by de Alfaro and Roy~\cite{dealfaro}. This variant did not compete last year, but entered the competition this year. This algorithm entered the competition as configuration $2$.

\subsubsection{Abstraction Refinement for Synthesis}
The idea of the abstraction-refinement algorithm is that, given an abstraction, we classify states into one of three categories: winning, losing, and unknown. If we discover that the entire initial set is winning, we know that the original game is winning and we can terminate. Dually, if we discover any initial state that is losing, we know that the entire initial set can never be winning, hence the game is losing and we can terminate. 

The algorithm iteratively refines the abstraction to reduce the number of states in the unknown classification until either:

\begin{itemize}
    \item all of the initial states are classified as winning (and the other states need not be classified, since this means that we can win from all initial states), or
    \item one of the initial states is classified as losing (again, no other states need to be classified)
\end{itemize}

The solver creates an abstraction by dropping a subset of state variables from the transition relation and instead allows them to be non-deterministically updated on each round of the game. The initial abstraction consists of only the variables that are mentioned in the safety specification, i.e., those that the error output $\out$ directly depends on.

\mysubsubsection{Abstract Game Solving}
The abstract game is based on two different interpretations of the transition relation, that compute the \emph{controllable predecessors} of a set of states with different assumptions: Cpre-must ($Cpre_1^M$) resolves the non-determinism of the abstraction in favor of the environment player, i.e., we assume that the system player can only force the game from an abstract state $s$ into an abstract state $s'$ if this is possible for \emph{any} valuation of the state variables that have been abstracted away. On the other hand, Cpre-may ($Cpre_1^m$) resolves the abstraction in favor of the system player, i.e., we assume that the system player can force the game from $s$ into $s'$ if there exists a valuation of the abstracted variables such that this is possible. 
%
%
%
%

Based on these functions, one can compute may-winning and must-winning regions, denoted $W^m$ and $W^M$ respectively, by iterating the functions above to a fixpoint.

\mysubsubsection{Abstraction Refinement}
Following de Alfaro and Roy~\cite{dealfaro}, we refine the abstraction at the may-must boundary, i.e. the boundary between states classified as winning and unknown. We split an abstract state into two such that from one of the new states $x$ is is possible for the environment player to force execution into the losing region. We find a candidate state to split by using an efficient symbolic calculation that determines the sub-states that are not must-winning but have a transition to the must-winning set $W^M$. These are the sub-states on the may-must boundary that are winning and are thus are part of a candidate state for splitting.

In order to extract a single sub-state on the may-must boundary that is winning, we extract a large prime implicant from this characteristic formula. This is an efficient operation with BDDs. We also extract the variables that occur in this prime that and are currently abstracted away.

Notice that if we include these variables in our abstraction, the abstraction is now precise enough to exactly represent the states in the implicant. These states are winning. This means that if we iterate the controllable predecessor one more time, we will at least discover that this implicant is winning, in addition to $W^M$. Thus, our refinement has grown the must winning set. In practice, refining the abstraction in this way usually discovers many more winning states than just the implicant. 
%
%

%
%
%
%
%
%
%

\subsubsection{Implementation, Availability}
Simple BDD solver is written in the Haskell functional programming language. It
uses the CUDD package for BDD manipulation and the
Attoparsec~\cite{attoparsec} Haskell package for fast parsing. Altogether, the
solver, AIGER parser, compiler and command line argument parser are just over
300 lines of code. The code is available online at:
\url{https://github.com/adamwalker/syntcomp}.

\section{Experimental Results}
\label{sec:results}

We present the results of \syntcomp 2015, separated into realizability and synthesis track. 
Detailed results of the competition are also directly accessible via the web-frontend of the EDACC platform at \url{syntcomp.cs.uni-saarland.de}.

\subsection{Realizability Track}

In the realizability track, tools competed on $250$ benchmark instances, selected from the different benchmark categories as explained in Section~\ref{sec:selection}. All $4$ tools that entered \syntcomp 2015 competed in the realizability track in at least one configuration. Overall, $10$ different configurations entered this track, with $7$ using sequential execution mode and $3$ using parallel mode. We first restrict the evaluation of results to purely sequential tools, then extend it to include also the parallel versions, and finally give a brief analysis of the results.

\paragraph{Sequential Mode.}
In sequential mode, all participants compete with at least one configuration: \abssynthe with three configurations (seq1, seq2 and seq3), \demiurge with one configuration (D1real), \realizer with one configuration (sequential), and \simpleBDD with two configurations (1 and 2).

The number of solved instances per configuration, as well as the number of uniquely solved instances, are given in Table~\ref{tab:results-realseq}. No tool could solve more than $200$ out of the $250$ instances, and $30$ instances could not be solved by any tool within the timeout. 

\begin{table}[h]
\caption{Results: Realizability (sequential mode only)}
\label{tab:results-realseq}
\centering
\def\arraystretch{1.2}
\begin{tabular}{c|ccc}
Tool (configuration)			& Solved  &  Unique \\ 
 \hline
\simpleBDD (2)             & 195 		& 9 \\
\abssynthe (seq2) 	 	 & 187 & 2 \\
\simpleBDD (1) & 185 & 0 \\
\abssynthe (seq3) & 179 & 0 \\
\realizer (sequential) & 179 & 0 \\
\abssynthe (seq1) & 173 & 1 \\
\demiurge (D1real) & 139 & 5 \\
\end{tabular}
\end{table}
The following benchmarks were solved uniquely by one tool configuration:
\begin{itemize}
\item \abssynthe (seq1): \texttt{moving\_obstacle\_128x128\_59glitches}
\item \abssynthe (seq2): \texttt{mult\_bool\_matrix\_6\_6\_6, mult12}
\item \demiurge (D1real): \texttt{beemldelec4b1\_c0to511, gb\_s2\_r2\_comp4\_REAL,\\ load\_full\_4\_comp1\_REAL, mult\_bool\_matrix\_dyn\_6\_6, very\_good\_bakery2.sym}
\item \simpleBDD (2): \texttt{amba9match5, cycle\_sched\_12\_6\_3, cycle\_sched\_8\_7\_2,\\ driver\_a7n, driver\_b7y, driver\_c8n, factory\_assembly\_5x5\_2\_10errors,\\ factory\_assembly\_5x5\_2\_11errors, good\_bakery.false}
\end{itemize}

\begin{figure}[h]
\centering
\includegraphics[width=\linewidth]{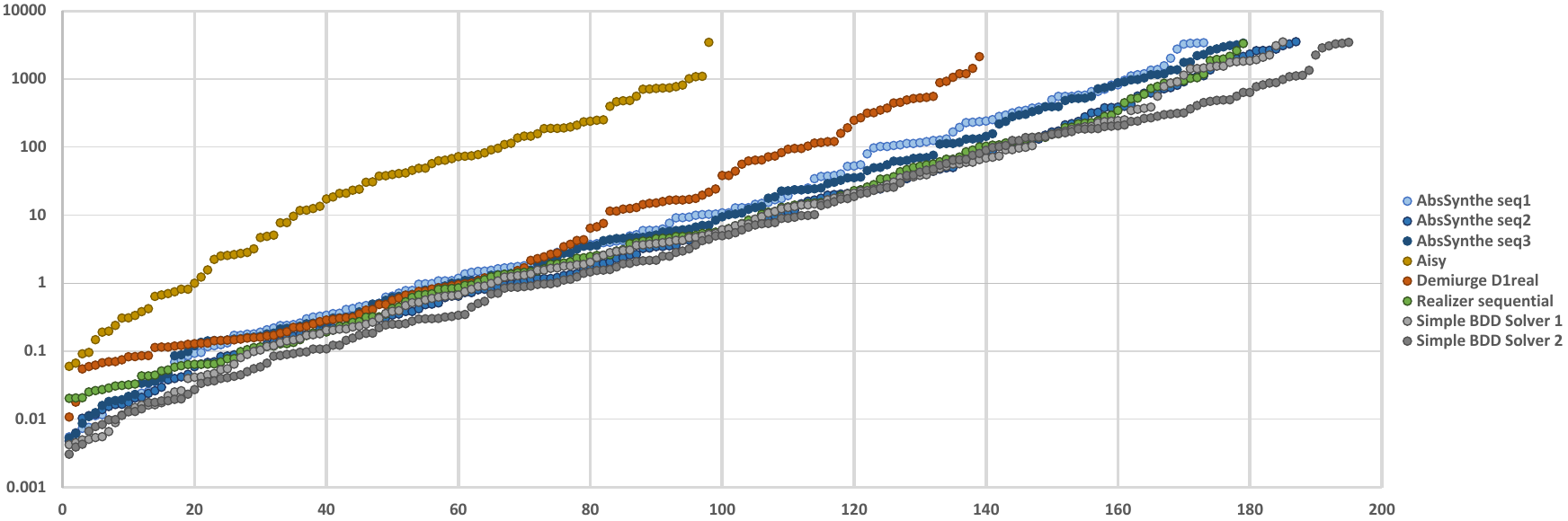}
\caption{Sequential Realizability Track, Runtime Cactus Plot}
\label{fig:cactus-realseq}
\end{figure}

For comparison, we also ran a number of additional tools on the benchmark set: last year's versions of \abssynthe, \simpleBDD, the learning-based sequential version of \demiurge, as well as Aisy~\cite{aisy}, our BDD-based, unoptimized reference implementation.\footnote{We don't consider last year's version of \realizer, since the improvements in the new version are limited to bug-fixes and the new parallel mode.} The results for these tools can be found in Table~\ref{tab:results-realseqref}. Furthermore, Figure~\ref{fig:cactus-realseq} gives a cactus plot for runtimes of sequential algorithms in the realizability track, including the reference implementation Aisy.

\begin{table}[h]
\caption{Results: Realizability (sequential, reference solvers)}
\label{tab:results-realseqref}
\def\arraystretch{1.2}
\centering
\begin{tabular}{c|ccc}
Tool (configuration)			& Solved \\ 
 \hline
\simpleBDD (2014)             & 182 \\
\abssynthe (2014) 	 	 & 169 \\
\demiurge (learn,2014) & 102 \\
Aisy & 98
\end{tabular}
\end{table}

\paragraph{Parallel Mode.}
Three of the tools that entered the competition had at least one parallel configuration for the realizability track: two configurations of \abssynthe (par1 and par2), and one configuration each of \demiurge (P3real) and \realizer (parallel). The difference to sequential mode is that the tools can use all four cores of the CPU, and the timeout is now measured in Wall Time instead of CPU Time. In particular, the parallel configurations had to solve the same set of benchmark instances as in the sequential mode. The results are given in Table~\ref{tab:results-realpar}. Again, no tool could solve more than $200$ instances, but a number of additional instances could be solved: only $11$ could not be solved by any tool in either sequential or parallel mode.

\begin{table}
\caption{Results: Realizability (parallel mode only)}
\label{tab:results-realpar}
\centering
\def\arraystretch{1.2}
\begin{tabular}{c|ccc}
Tool (configuration)			& Solved  &  Unique \\ 
 \hline
\abssynthe (par1) 	 	 & 193 & 0 \\
\realizer (parallel) & 185 & 3 \\
\demiurge (P3real) & 183 & 17 \\
\abssynthe (par2) & 173 & 0 \\
\end{tabular}
\end{table}

\begin{figure}[h]
\centering
\includegraphics[width=\linewidth]{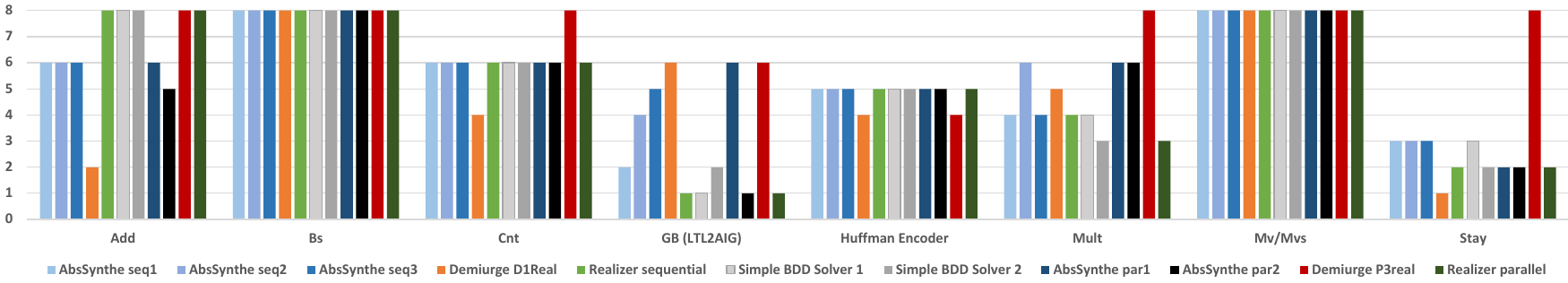}
\caption{Realizability Track, Solved Instances by Category}
\label{fig:bycat1}
\end{figure}

\begin{figure}[h]
\centering
\includegraphics[width=\linewidth]{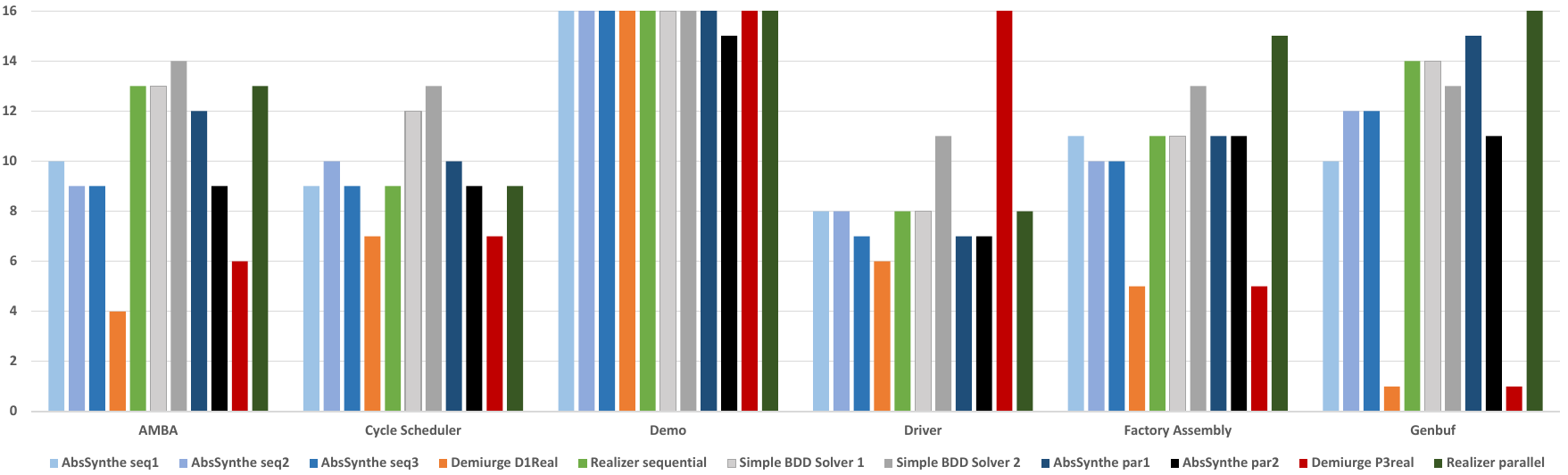}
\caption{Realizability Track, Solved Instances by Category}
\label{fig:bycat2}
\end{figure}

\begin{figure}[h]
\centering
\includegraphics[width=\linewidth]{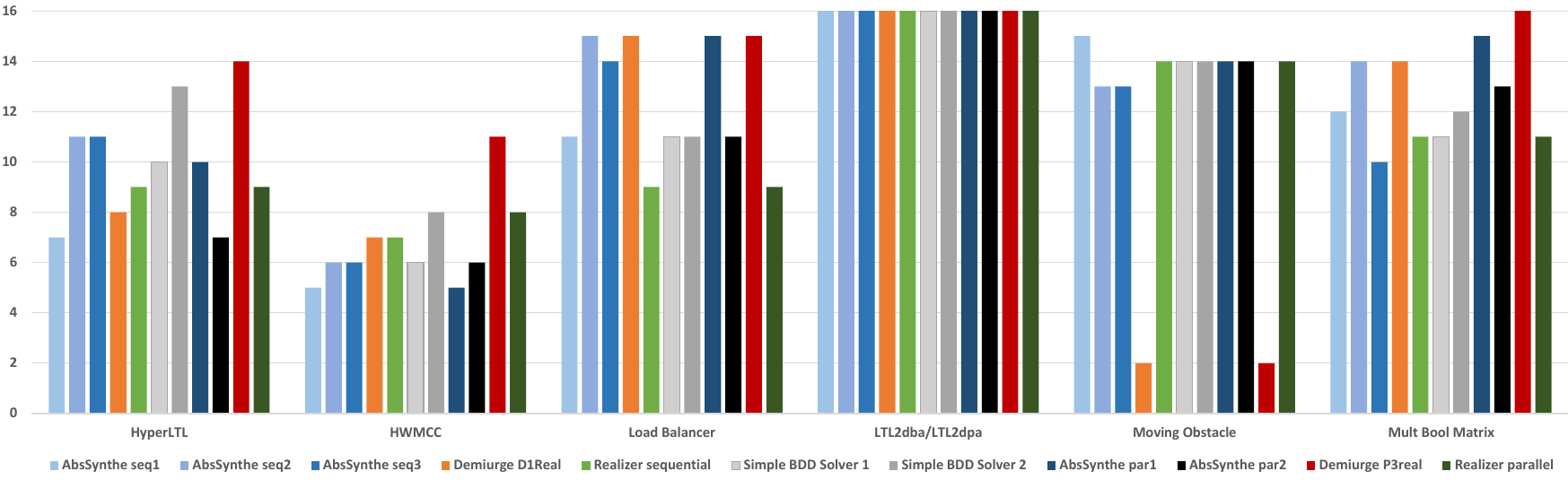}
\caption{Realizability Track, Solved Instances by Category}
\label{fig:bycat3}
\end{figure}

For an analysis of benchmark instances by category, Figures~\ref{fig:bycat1}, \ref{fig:bycat2} and \ref{fig:bycat3} give an overview of the number of solved instances per configuration and category.

Note also that in Table~\ref{tab:results-realpar} we only count a benchmark instance as uniquely solved if it is not solved by any other configuration, including the sequential configurations.
Considering both sequential and parallel configurations, the following benchmarks were solved uniquely by one tool configuration:
\begin{itemize}
\item \demiurge (P3real): \texttt{6s216rb0\_c0to31.aag, cnt30n.aag, cnt30y.aag, driver\_a10n.aag, driver\_a8n.aag, driver\_b10y.aag, driver\_b8y.aag, driver\_c10n.aag, mult14.aag, mult16.aag, oski2ub1i\_c0to7.aag, oski3ub1i\_c0to255.aag, stay18y.aag,\\ stay20n.aag, stay20y.aag, stay22n.aag, stay22y.aag},
\item \realizer (parallel): \texttt{factory\_assembly\_7x5\_2\_10errors.aag,\\ factory\_assembly\_7x5\_2\_11errors.aag, genbuf64c2unrealy.aag},
\item \simpleBDD (2): \texttt{cycle\_sched\_12\_6\_3.aag, cycle\_sched\_8\_7\_2.aag}.
\end{itemize}

Considering again last year's versions, only \demiurge ran in a proper parallel mode in 2014.\footnote{\abssynthe and \realizer had parallel modes, but due to bugs and faulty call parameters did not produce interesting results.} Out of the $250$ benchmark instances, \demiurge (parallel, 2014) could solve $99$ instances. This is slightly less than the sequential configuration from 2014, as is to be expected based on last year's results~\cite{Jacobs15}.

\begin{figure}[h]
\centering
\includegraphics[width=\linewidth]{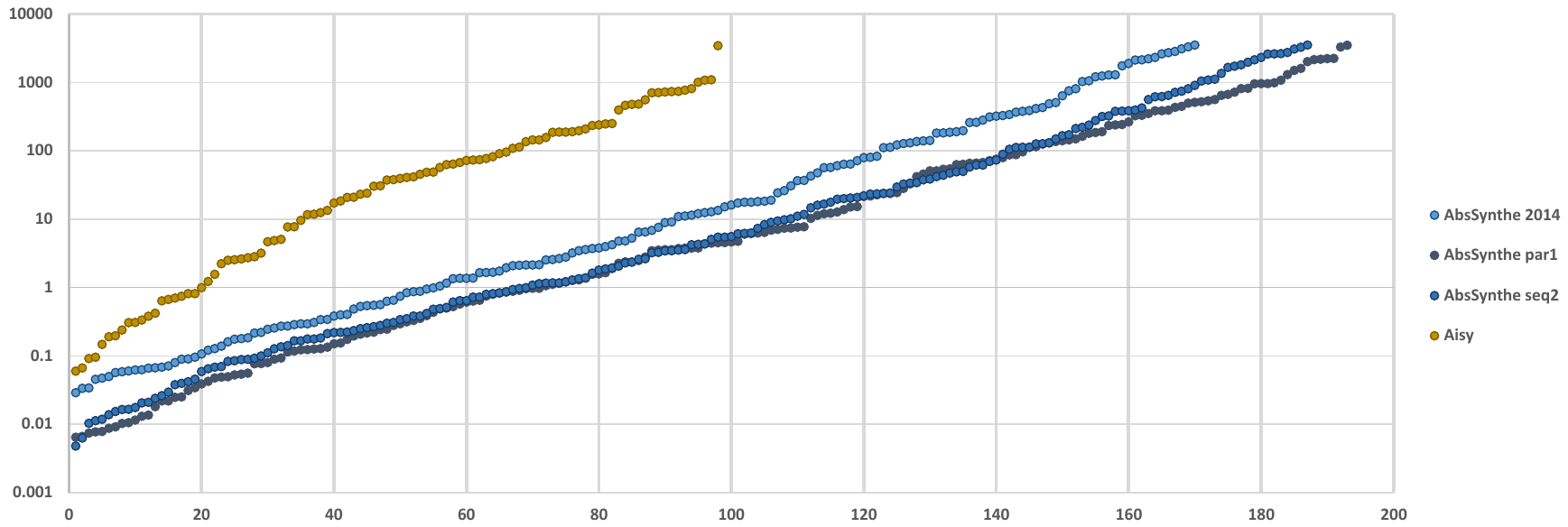}
\caption{Realizability Track, Improvement of \abssynthe over Last Year}
\label{fig:cactus-abssynthe}
\end{figure}

\begin{figure}[h]
\centering
\includegraphics[width=\linewidth]{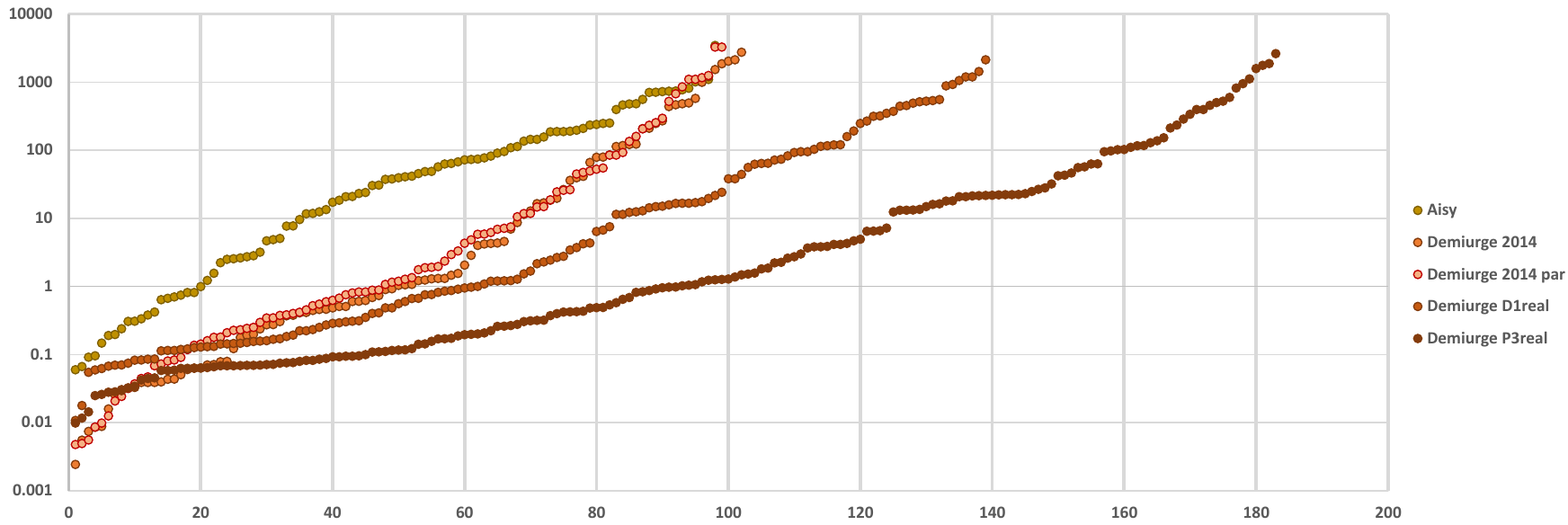}
\caption{Realizability Track, Improvement of \demiurge over Last Year}
\label{fig:cactus-demiurge}
\end{figure}

\begin{figure}[h]
\centering
\includegraphics[width=\linewidth]{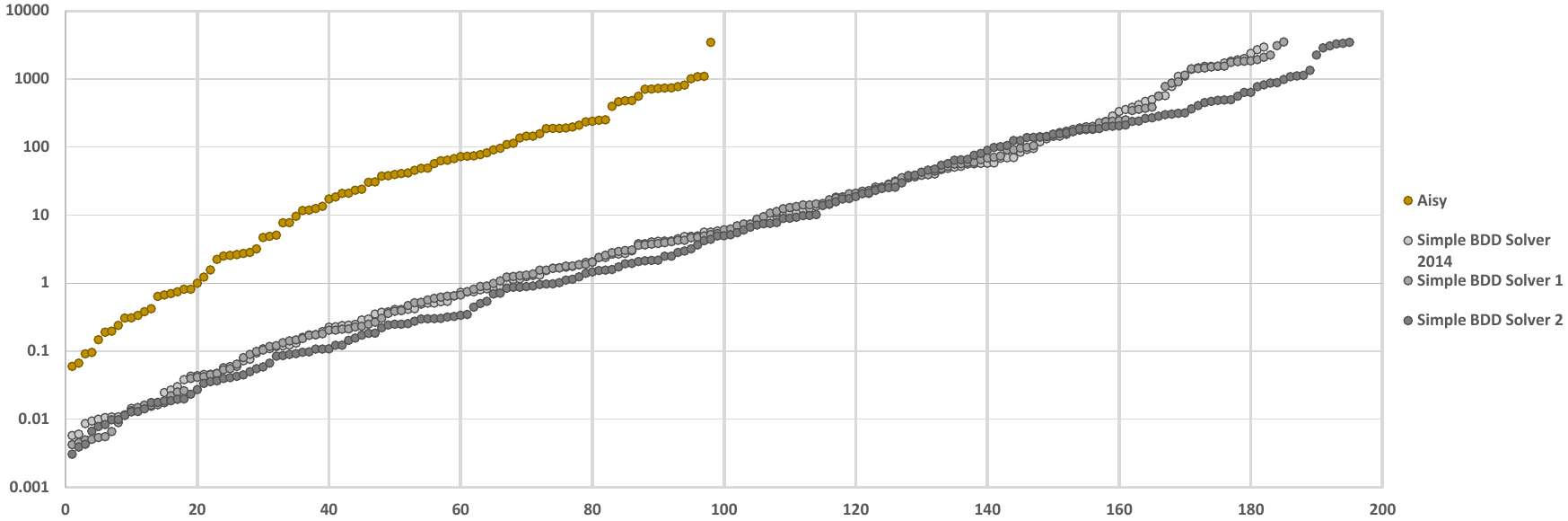}
\caption{Realizability Track, Improvement of \simpleBDD over Last Year}
\label{fig:cactus-simpleBDD}
\end{figure}

Finally, to further investigate the improvement of \abssynthe, \demiurge, and \simpleBDD compared to last year's versions, we give additional cactus plots that compare the best-performing configurations from each tool from 2014 and 2015 in Figures~\ref{fig:cactus-abssynthe}, \ref{fig:cactus-demiurge}, and \ref{fig:cactus-simpleBDD}.

\paragraph{Analysis.}
We note that on this year's benchmark set, the sequential configuration \abssynthe (seq2) solve slightly more problems than last year's winner, represented by configuration \simpleBDD (1). However, the new abstraction-based configuration \simpleBDD (2) can solve $10$ additional problems, which is slightly more than the best parallel version of \abssynthe, configuration (par1).

Comparing the participants to the unoptimized reference implementation Aisy, we see in Figure~\ref{fig:cactus-realseq} that the BDD-based participants can be considered to be two orders of magnitude faster than Aisy, and in Figure~\ref{fig:cactus-demiurge} that \demiurge (P3real) is faster by three orders of magnitude on a lot of problems.

Considering the different configurations of \abssynthe, we note that the best-performing sequential configuration is (seq2), which uses the compositional algorithm with global aggregation. Configuration (seq1), which is essentially the same as last year's version, solves the least number of problems. The best configuration overall is the parallel configuration (par1), which runs the different compositional algorithms as well as the non-compositional algorithm (seq1) in parallel. Running multiple threads of (seq1) in parallel, only with different BDD reordering strategies, did not pay off: configuration (par2) even solves less problems than (seq1).

Finally, \demiurge shows an impressive improvement over last year's version: the sequential configuration \demiurge (D1real) can solve $37$ additional instances when compared to last year's sequential (learn) configuration, and the new parallel mode solves $44$ more problems than the sequential configuration, or $86$ more problems than last year's parallel configuration. Comparing it to the other tools on different benchmark classes (see Figures~\ref{fig:bycat1}, \ref{fig:bycat2} and \ref{fig:bycat3}), we note that it still cannot compete with the other tools on benchmark classes such as AMBA, Factory Assembly, Genbuf, and Moving Obstacle, but it outperforms all other tools on benchmark classes Cnt, Mult, Stay, Driver, HyperLTL, HWMCC, and Mult Bool Matrix, most of which are new benchmarks that have been added this year.

\subsection{Synthesis Track}
In the synthesis track, tools competed on the same benchmarks as in the realizability track, except that those instances that could not be solved by any configuration (sequential or parallel) in the realizability track have been removed from the benchmark set. Thus, the benchmark set for the synthesis track contains $239$ instances.
Two tools entered the synthesis track: \abssynthe with three sequential and two parallel configurations, and \demiurge with one configuration for each mode.

In the synthesis track, we have two different ranking schemes: the basic ranking is by number of problem instances that can be solved within the timeout, and the separate \emph{quality ranking} gives more points to small solutions of realizable specifications, as explained in Section~\ref{sec:rules}. In both cases, a solution for a realizable specification is only considered as correct if it can be model-checked within a separate timeout of one hour (cf. Sections~\ref{sec:rules} and \ref{sec:execution}). Like in the realizability track, in the following we start by presenting the results for the sequential configurations, followed by parallel configurations, and end with an analysis of the results.

\paragraph{Sequential Mode.}
In this mode, \abssynthe competed with three configurations (seq1, seq2, seq3), and \demiurge with one configuration (D1synt). In addition, we ran last year's versions of \abssynthe and the learning-based configuration of \demiurge, as well as our reference implementation Aisy, on this year's competition benchmarks. 

Table~\ref{tab:results-syntseq} summarizes the experimental results, including the number of solved benchmarks, the number of points in the quality ranking, the uniquely solved instances, and the number of solutions that could not be model-checked within the timeout. Note that in the table a (realizable) problem instance is only considered as solved if the tool presents a solution that can be model-checked. With this requirement, no tool could solve more than $161$ or about $67\%$ of the benchmarks, and $60$ instances could not be solved by any tool. In particular, for all configurations of \abssynthe there is a rather high number of benchmarks for which the tool can provide a solution, but this solution cannot be model-checked.


\begin{table}[h]
\caption{Results: Synthesis (sequential mode only)}
\label{tab:results-syntseq}
\centering
\def\arraystretch{1.2}
\begin{tabular}{c|cccc}
Tool (configuration) & Solved & Quality & Unique & MC Timeout\\
\hline
\abssynthe (seq2) & 161 & 254 & 4 & 16\\
\abssynthe (seq3) & 152 & 241 & 1 & 16\\
\abssynthe (seq1) & 148 & 234 & 6 & 18\\
\abssynthe (2014) & 145 & 231 & \% & 16\\
\demiurge (D1synt) & 127 & 214 & 8 & 4\\
\demiurge (2014,learn) & 83 & 138 & \% & 1\\
Aisy & 75 & 105 & \% & 3
\end{tabular}
\end{table}

\paragraph{Parallel Mode.}
In this mode, \abssynthe competed with two configurations (par1, par2), and \demiurge with one configuration (P3synt). In addition, we ran last year's parallel configuration of \demiurge on the new competition benchmarks.

Table~\ref{tab:results-syntpar} summarizes the experimental results, again including the number of solved benchmarks, the number of points in the quality ranking, the uniquely solved instances, and the number of solutions that could not be model-checked within the timeout. No tool could solve more than $180$ problem instances, or about $75\%$ of the benchmark set. The number of (potential) solutions that cannot be model checked within the timeout is again rather high for \abssynthe, while \demiurge only produces a single solution that cannot be checked.
Like in the parallel realizability track, we only consider instances as uniquely solved if they are not solved by any other configuration, including sequential ones.

\begin{figure}[h]
\centering
\includegraphics[width=\linewidth]{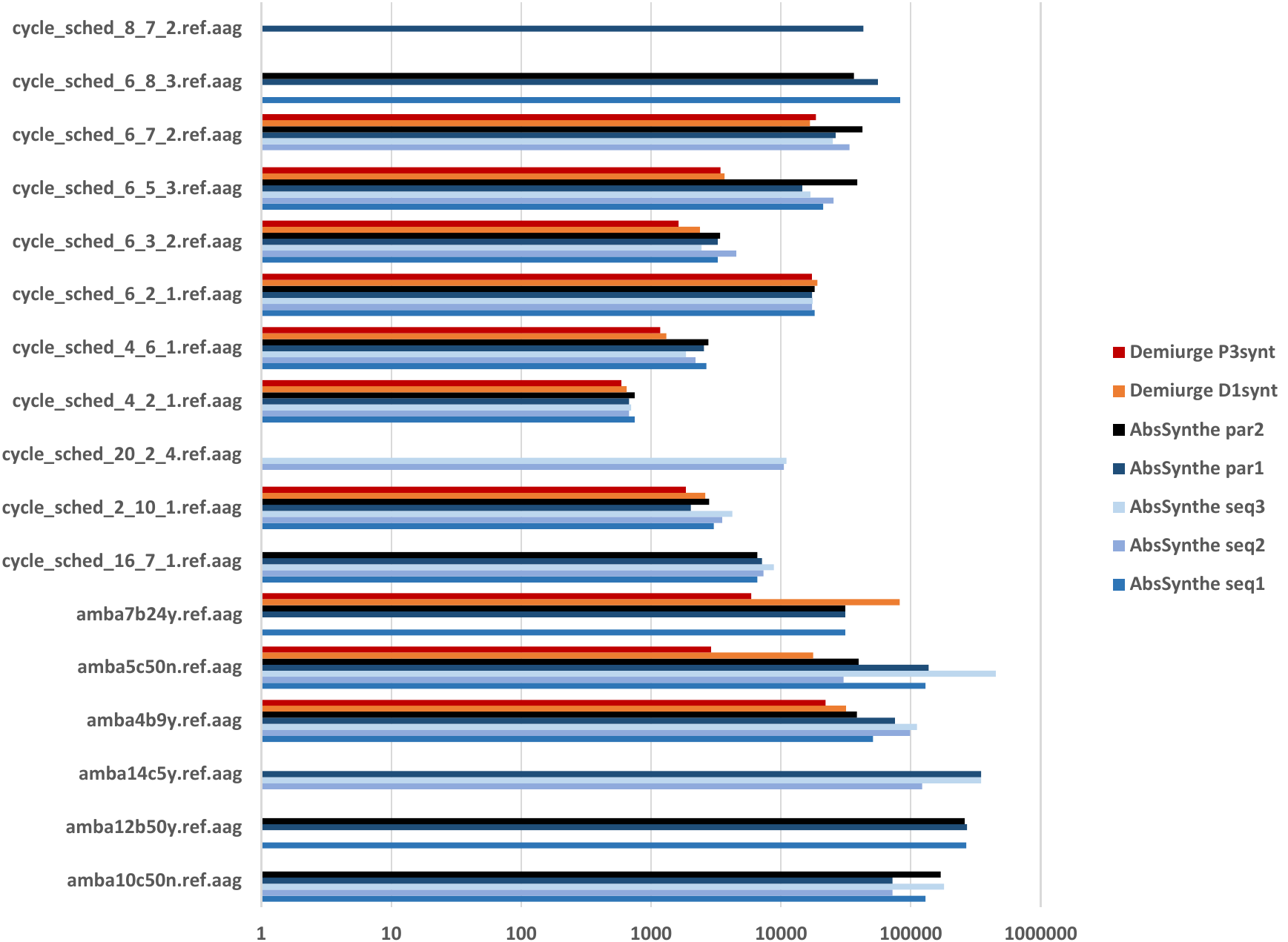}
\caption{Comparison of Solution Size for AMBA and Cycle Scheduling Benchmarks}
\label{fig:size1}
\end{figure}
\begin{figure}[h]
\centering
\includegraphics[width=\linewidth]{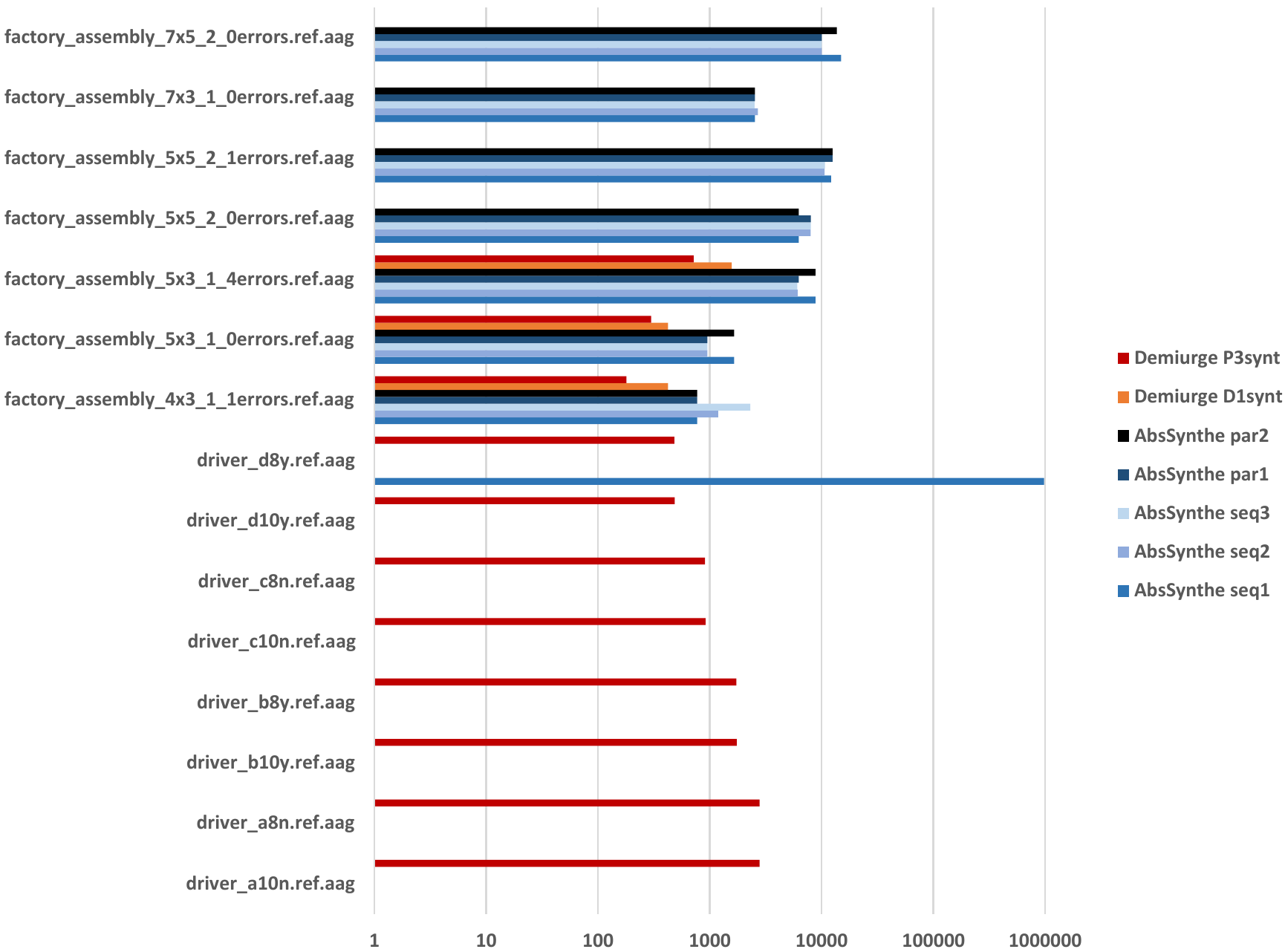}
\caption{Comparison of Solution Size for AMBA and Cycle Scheduling Benchmarks}
\label{fig:size2}
\end{figure}
\begin{figure}[h]
\centering
\includegraphics[width=\linewidth]{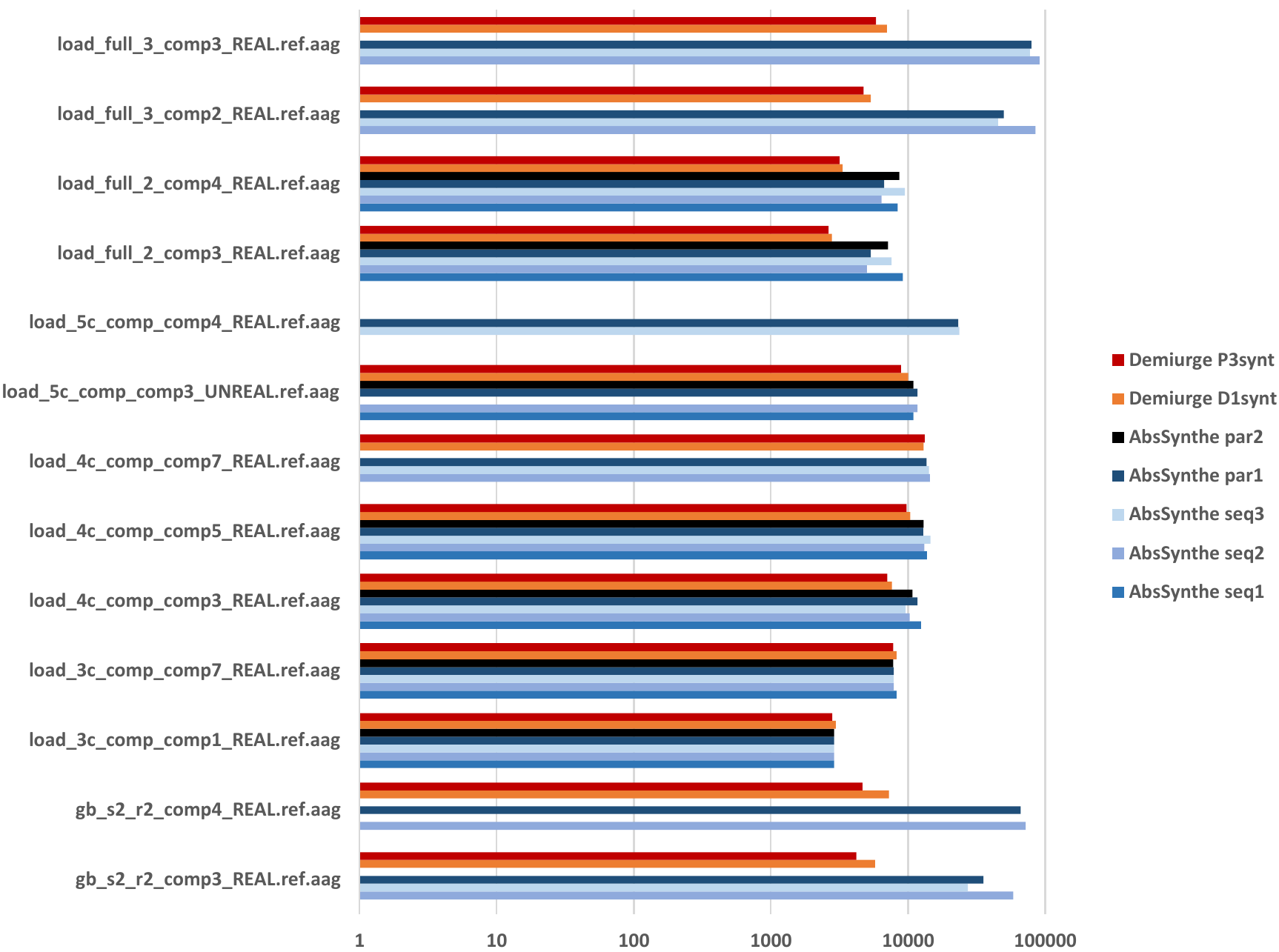}
\caption{Comparison of Solution Size for AMBA and Cycle Scheduling Benchmarks}
\label{fig:size3}
\end{figure}

\begin{table}[h]
\caption{Results: Synthesis (parallel mode only)}
\label{tab:results-syntpar}
\centering
\def\arraystretch{1.2}
\begin{tabular}{c|cccc}
Tool (configuration) & Solved & Quality & Unique & MC Timeout\\
\hline
\demiurge (P3Synt) & 180 & 317 & 28 & 1\\
\abssynthe (par1) & 167 & 263 & 2 & 20\\
\abssynthe (par2) & 148 & 235 & 0 & 17\\
\demiurge (2014,parallel) & 88 & 144 & 0 & 1
\end{tabular}
\end{table}
\paragraph{Analysis.}
Consider the fact that all configurations of \abssynthe produce a high number of (potential) solutions that cannot be model checked. In contrast, \demiurge only produces very few such solutions. We assume that this is in part due to the fact that \demiurge in most cases produces smaller solutions than \abssynthe. We compare solution sizes for some problem instances in Figures~\ref{fig:size1}, \ref{fig:size2} and \ref{fig:size3}.


Apart from the issue of non-verifiable solutions, the performance of the different algorithms is exactly what is to be expected, based on the results of the realizability track.
In particular, note that without the requirement of verifiable solutions, the parallel configuration \abssynthe (par1) would have ranked higher than \demiurge (P3synt).\footnote{We tried to resolve the problem of model checking timeouts by running other model checkers on a number of the solutions that cannot be verified. In particular, we tried the other two highest-ranking tools for single safety properties in HWMCC 2014, \Abc~\cite{abc} and V3~\cite{V3,WuWLH14} However, neither of the tools was successful on the solutions that we tried to verify.}

Also note that, although our quality metric does not change the ranking of tools, it does increase the relative difference in some cases. E.g., the difference between \demiurge (P3synt) and \abssynthe (par1) is only $8\%$ in number of solved instances, but $20\%$ in the quality score. An analysis of the size of the potential solutions that could not be model checked shows that none of them is smaller than the reference size, which implies that \abssynthe (par1) would be ranked below \demiurge (P3synt) in the quality ranking, even if all of its solutions could be model checked.

Considering uniquely solved instances, we note that the differences are rather small when we only compare the sequential configurations --- each configuration produces some unique solutions, and \abssynthe (seq1) produces almost as many as \demiurge (D1synt). This picture changes when we look at the parallel configurations: \demiurge (P3synt) has a high number of uniquely solved instances, i.e., instances that can neither be solved by any sequential configuration, nor by the parallel configurations of \abssynthe. In contrast, almost all instances that are solved by \abssynthe (par1) and \abssynthe (par2) are also solved by the sequential configurations, or by \demiurge (P3synt).

\section{Conclusions}
\label{sec:conclusions}

\syntcomp 2015 was another big step towards establishing the competition in the synthesis community and extending the benchmark format and library. We have collected thousands of additional benchmark instances, and refined our ranking and evaluation for a fairer comparison of participants. In addition, our experimental results show that some of the tools have made impressive improvements when compared to last year's versions.

For 2016, we will consider an extension of the problem from
pure safety specifications to a specification format that includes liveness,
possibly in the form of LTL formulas.

\myparagraph{Acknowledgements}
We thank Armin Biere and R\"udiger Ehlers for advice on the changes to the competition rules, and Ayrat Khalimov
for supplying the reference implementation Aisy~\cite{aisy}, as well as the Huffman Encoder Benchmarks. We also thank Markus Rabe for supplying the HyperLTL benchmarks. Finally, we thank Jens Kreber for technical assistance during setup and execution of the competition at Saarland University.

The organization of \syntcomp 2014 was supported by the Austrian Science Fund
(FWF) through projects RiSE (S11406-N23) and QUAINT (I774-N23), by the German
Research Foundation (DFG) as part of the Transregional Collaborative Research
Center ``Automatic Verification and Analysis of Complex Systems'' (SFB/TR 14
AVACS) and through project ``Automatic Synthesis of Distributed and
Parameterized Systems'' (JA 2357/2-1).

The development of \abssynthe was supported by an F.R.S.-FNRS fellowship, and
the ERC inVEST (279499) project.

The development of \demiurge was supported by the FWF through projects RiSE
(S11406-N23, S11408-N23) and QUAINT (I774-N23).

The development of \realizer was supported by the DFG as part of SFB/TR 14
AVACS.

The development of \simpleBDD was supported by a gift from the Intel
Corporation, and NICTA is funded by the Australian Government through the
Department of Communications and the Australian Research Council through the ICT
Centre of Excellence Program.

\bibliographystyle{eptcs}
\bibliography{synthesis}
\end{document}